\documentclass[fleqn,10pt,twocolumn]{wlscirep}
\usepackage[normalem]{ulem}
\usepackage[utf8]{inputenc}
\usepackage[T1]{fontenc}
\usepackage{ragged2e}
\usepackage{dcolumn}
\usepackage{amsmath}
\usepackage{epsfig}
\usepackage{float}
\usepackage{graphicx}
\usepackage{subfigure}
\usepackage{hyperref}
\usepackage{colortbl}
\usepackage{xcolor}
\usepackage{multirow}
\usepackage{amsthm,amsmath,amssymb}
\usepackage{mathrsfs}
\usepackage{bm}
\graphicspath{{./images/}}
\newcommand{\jj}{\mathrm{j}}

\title{Ultra-Large-Capacity Passive Quantum Access Network Powered By Single Thermal Source}

\author[1]{Yuehan Xu}
\author[1]{Qijun Zhang}
\author[1]{Xiaojuan Liao}
\author[1]{Zidong Gao}
\author[1]{Piao Tan}
\author[1]{Xufeng Liang}
\author[1]{Hanwen Yin}
\author[1,2,3]{Peng Huang}
\author[1,2,3,*]{Tao Wang}
\author[1,2,3,$\dagger$]{Guihua Zeng}

\affil[1]{State Key Laboratory of Photonics and Communications, Center of Quantum Sensing and Information Processing, Shanghai Jiao Tong University, Shanghai 200240, China}
\affil[2]{Shanghai Research Center for Quantum Sciences, Shanghai 201315, China}
\affil[3]{Hefei National Laboratory, Hefei 230088, China}
\affil[*]{tonystar@sjtu.edu.cn}
\affil[$\dagger$]{ghzeng@sjtu.edu.cn}

\begin{abstract}
    
	Quantum Key Distribution (QKD) aims to provide secure keys for classical communications through one-time pad (OTP) encryption, guaranteed by physical laws. Currently, an advanced and dominant Classical Access Network (CAN) based on Passive Optical Network (PON) can support up to $256$ $\mathrm{users}$ with a total rate of $10$ $\mathrm{Gbps}$ (denoted as $10\mathrm{-Gbps}$ $@$ $256\mathrm{-users}$). The equivalent rate requirement brought by OTP encryption makes it necessary for the performance of the QKD Access Network (QAN) to be maintained at the same level as CAN. However, state-of-the-art QAN based on PON is far from meeting this standard. To address this issue, we propose a passive QAN that distributes polychromatic quantum randomness from a single thermal source, supporting up to $304$ $\mathrm{users}$ with an aggregate secret key rate (SKR) of $13$ $\mathrm{Gbps}$ (denoted as $13\mathrm{-Gbps}$ $@$ $304\mathrm{-users}$). The excellent performance of the Thermal State QAN (TS-QAN) comes from three aspects: The thermal states, with their broad spectrum and Bose-Einstein distribution, can be represented, through the Glauber--Sudarshan representation, as high-bandwidth Gaussian coherent-state ensembles across different frequency modes, eliminating the need for numerous active modulation devices and Quantum Random Number Generators (QRNGs) in a polychromatic network; The beacons enabled by an Electro-Optic (EO) comb achieve time-varying and polychromatic phase tracking, where each frequency-mode thermal signal can be coherently measured with a Local Local Oscillator (LLO) aided by the corresponding beacon, removing the requirement for large-scale phase-locking networks; The state broadcasting technology allows each user to obtain independent final keys via reverse reconciliation after residual broadcast-induced correlations are accounted for, expanding large network capacity with small SKR losses. Ultimately, the experiment verified the feasibility of $13\mathrm{-Gbps}$ $@$ $304\mathrm{-users}$ TS-QAN using Continuous-Variable QKD (CV-QKD) under a covariance-matrix-based network security analysis including multimode Holevo leakage and residual broadcast correlations. This work meets the demands for SKR and network capacity from advanced and dominant CAN to QAN ($13\mathrm{-Gbps}$ $@$ $304\mathrm{-users}$ meeting $10\mathrm{-Gbps}$ $@$ $256\mathrm{-users}$), providing a scalable and high-performance solution for deploying QAN in modern telecommunication systems.
    
\end{abstract}

\begin{document}
\flushbottom
\maketitle
\thispagestyle{empty}

\section*{Introduction}

Due to the impact of quantum computers \cite{arute2019quantum}, the security of the Rivest-Shamir-Adleman (RSA) public key algorithm \cite{rivest1978method} used in classical communication is increasingly precarious. Quantum Key Distribution (QKD) \cite{scarani2009security}, guaranteed by the physical laws, is one of the best solutions to this problem. QKD achieves real-time encryption of classical information through physical links and one-time pad (OTP) encryption, reaching information-theoretic unconditional security. OTP requires that the length of the classical information matches that of the quantum key; that is, their rates must be comparable. This requirement has driven much research on point-to-point QKD to focus on improving the secret key rate (SKR) to match modern communication rates \cite{wang2022sub, grunenfelder2023fast, li2023high, hajomer2024cv, ji2024gbps, wang2025high, hajomer2025chip}. While point-to-point QKD has been advancing rapidly \cite{grosshans2002continuous, wang2005beating, peng2007experimental, bennett2014quantum, qi2015generating, huang2016long, liao2017satellite, lucamarini2018overcoming, yin2020entanglement, zhang2020long, wang2022twin, liu2023experimental, hajomer2024long}, QKD networks that have been emerging in recent years have also brought forth various excellent networking solutions \cite{townsend1997quantum, choi2011quantum, frohlich2013quantum, frohlich2015quantum, wengerowsky2018entanglement, dynes2019cambridge, joshi2020trusted, chen2021implementation, chen2021integrated, huang2021realizing, wang2021practical, fan2022robust, mandil2023quantum, wang2023experimental, xu2023round, hajomer2024continuous, huang2024cost, li2024experimental, liu2024integrated, qi2024experimental, xu2024integrated, pan2025high, xu2025ofdm, xu2026polychromatic}. However, as the hardware complexity scaling, network capacity, and communication rate of classical communication networks continue to increase, the current development of QKD networks is far from meeting the OTP encryption demands of modern communication networks.

This paper aims to address the performance gap issue between QKD networks and classical communication networks deployed in optical access networks, which need to meet the data transmission and encryption demands of many users in the last mile. The optical network architecture can be categorized into Active Optical Networks (AONs) and Passive Optical Networks (PONs). The main distinction lies in the optical distribution components. AON utilizes active optical components, such as optical switches, for light distribution. In contrast, PON employs passive optical components like Beam Splitters (BSs), Wavelength Division Multiplexers (WDMs), and Dense Wavelength Division Multiplexers (DWDMs) for the same purpose. In optical access networks, communication occurs between one Optical Line Terminal (OLT) and multiple Optical Network Units (ONUs). This point-to-multipoint structure relies on an Optical Distribution Network (ODN) for light distribution. Using passive devices as ODN, PON offers a low-cost, large-capacity, and high-speed solution, making it more suitable for user access scenarios than AON. Consequently, PON has been widely deployed in global access networks. For classical communication, the performance of advanced and dominant Classical Access Networks (CANs) based on PON structures can reach $10\mathrm{-Gbps}$ $@$ $256\mathrm{-users}$ \cite{kramer2002ethernet, banerjee2005wavelength, mcgarry2006wdm, lam2011passive, abbas2016next, wey2018passive}. For QKD, the state-of-the-art QKD Access Networks (QANs) based on PON structures primarily utilize DWDMs and BSs, with their optimal performance being $8.75\mathrm{-Gbps}$ $@$ $19\mathrm{-users}$ \cite{xu2026polychromatic} and $16.80\mathrm{-Mbps}$ $@$ $8\mathrm{-users}$ \cite{hajomer2024continuous} / $33.38\mathrm{-Mbps}$ $@$ $16\mathrm{-users}$ \cite{pan2025high}, respectively. Apart from the well-known hardware complexity scaling gap, the performance of QAN also falls far behind that of CAN. This situation hinders the widespread application of QKD in households.

\begin{figure*}[!h]
	\centering
	\subfigure[]{\label{fig1a}\includegraphics[height=12.33cm]{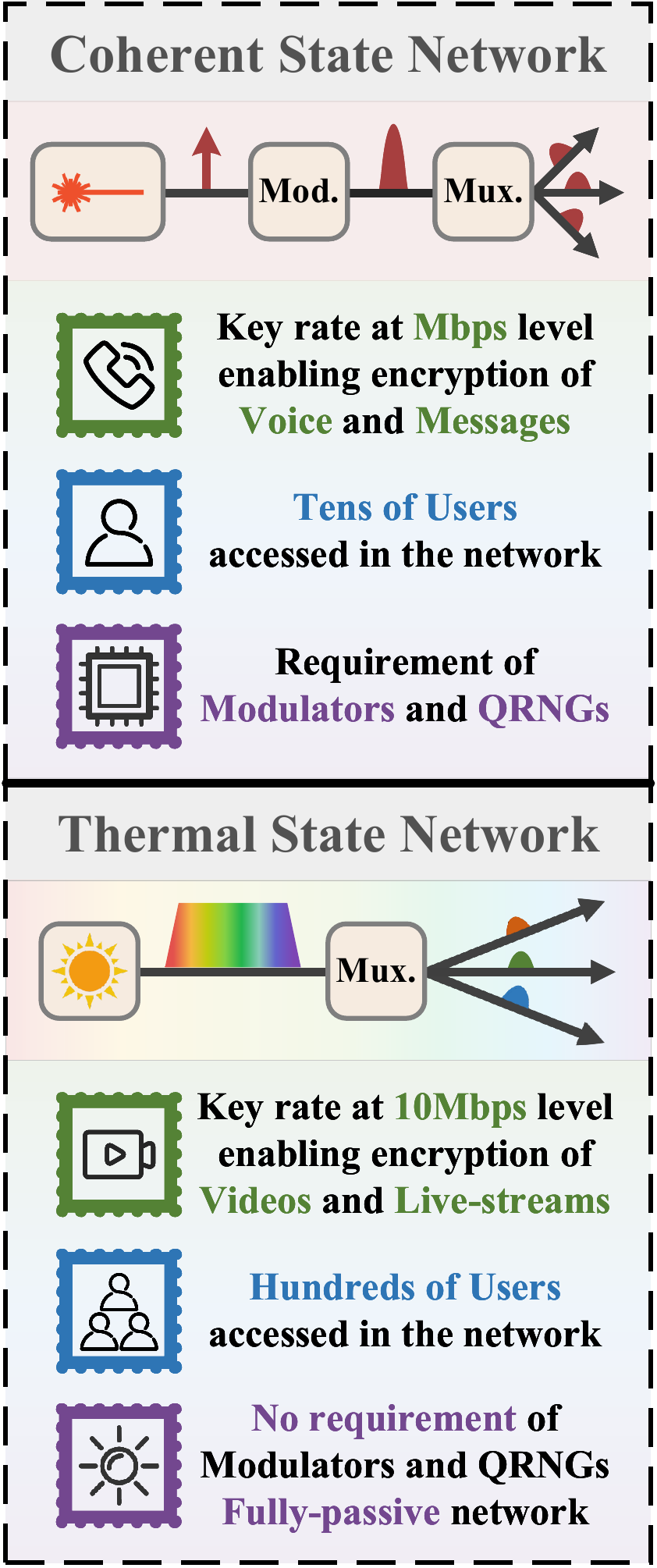}}
	\subfigure[]{\label{fig1b}\includegraphics[height=12.33cm]{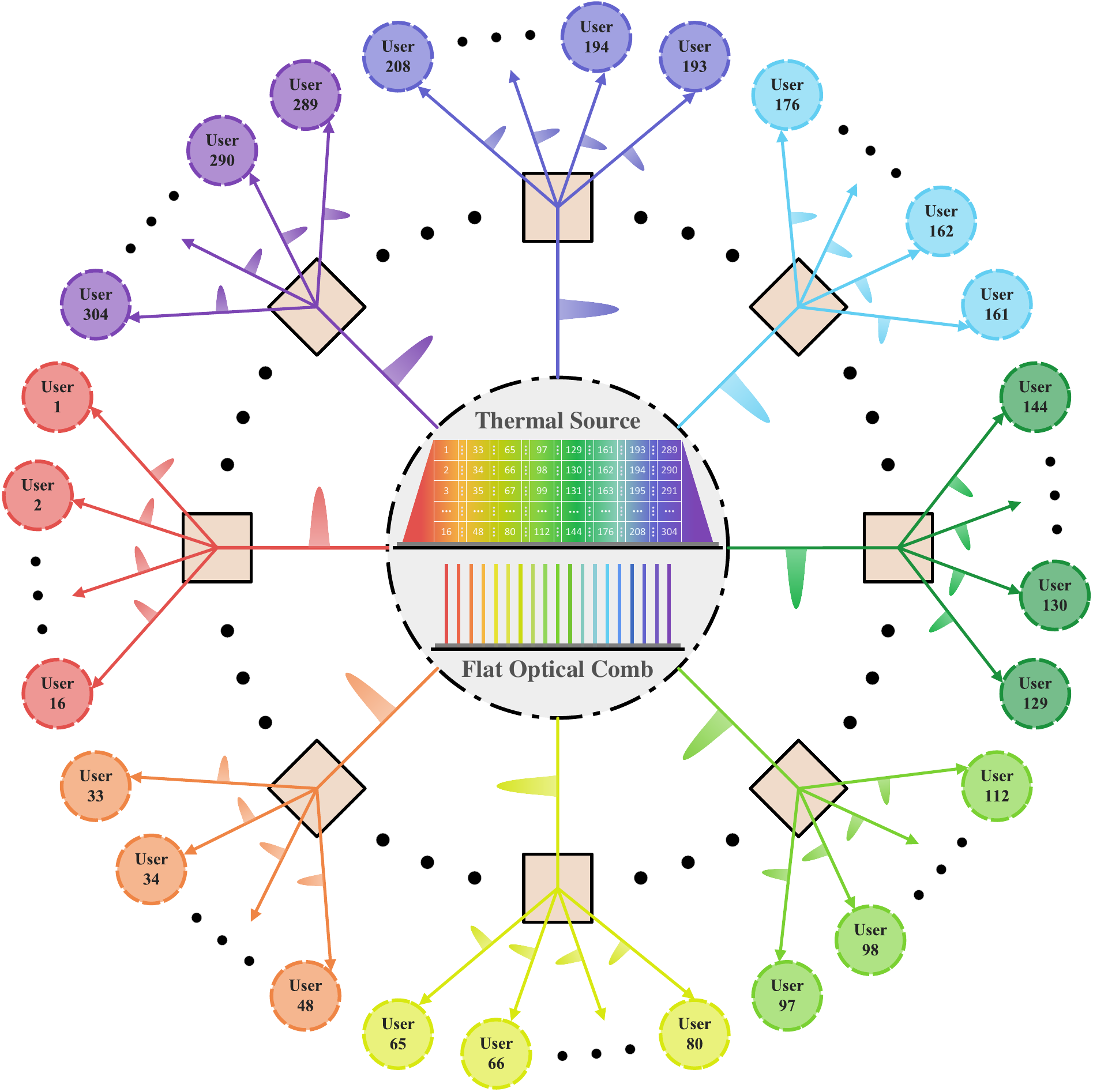}}
	\caption{Brief introduction to the TS-QAN. (a) Comparison between the coherent state network and the thermal state network. Mod. and Mux. correspond to the modulator and multiplexer, respectively. It is intended as an architecture-level illustration based on representative reported access-network demonstrations, rather than as a controlled comparison under matched hardware and symbol-rate conditions. (b) Implementation and topology of the thermal state network. With the thermal source and flat optical comb as the core, the thermal state network achieves point-to-multipoint QKD for $304$ $\mathrm{users}$.}
	\label{fig1}
\end{figure*}

Let's discuss the challenges of QAN from three perspectives: hardware complexity scaling, network capacity, and SKR. DWDM-based QAN \cite{huang2024cost,xu2026polychromatic} utilizes polychromatic resources to distribute quantum states, achieving a notable SKR per user. However, it requires preparation of polychromatic resources, multiple active modulation devices, and multiple Quantum Random Number Generators (QRNGs), which significantly increase the cost and limit the scalability of network capacity. On the contrary, BS-based QAN \cite{townsend1997quantum, choi2011quantum, frohlich2013quantum, frohlich2015quantum, huang2021realizing, wang2021practical, wang2023experimental, xu2023round, hajomer2024continuous, li2024experimental, qi2024experimental, pan2025high, xu2025ofdm} uses monochromatic beam splitting to distribute quantum states, achieving a favorable hardware complexity scaling. However, its limitation lies in the high loss brought by beam splitting, which causes a sharp decline in the SKR as network capacity increases. While active CV-QKD with multi-GHz repetition rate has been achieved in point-to-point links \cite{hajomer2024cv, wang2025high, hajomer2025chip}, scaling active state preparation to a multi-user quantum access network typically introduces additional network-level constraints, including parallel high-speed chain replication, synchronization or calibration overhead, and sophisticated Digital Signal Processing (DSP). As a result, current reported access-network demonstrations are typically in the sub-GHz regime \cite{hajomer2024continuous, pan2025high, xu2025ofdm}. Essentially, the finite network-level utilization of high-bandwidth quantum randomness further limits the performance of QAN.

To address these issues, we propose the Thermal State QAN (TS-QAN). It integrates the passive preparation structure \cite{qi2018passive, qi2020experimental, huang2021experimental, wu2021passive, zhang2023experimental, ji2024gbps, yin2025all}, the polychromatic distribution scheme \cite{xu2026polychromatic}, and the state broadcasting method \cite{hajomer2024continuous, pan2025high}, providing a passive QAN solution that is capable of matching the performance of CAN. First is hardware complexity scaling. The broadband thermal state can be represented, through the Glauber--Sudarshan representation and continuous-mode quantum optics, as Gaussian coherent-state ensembles across multiple frequency modes \cite{raymer1989temporal, blow1990continuum, fabre2020modes, raymer2020temporal, chen2023continuous, sun2025analyzing}, replacing the preparation of polychromatic resources. The passive preparation structures, although they require high-performance detection and digitization, produce quantum states available for distribution, obviating the need for multiple active modulation devices and multiple QRNGs. Polychromatic beacon light based on the Electro-Optic (EO) comb and time-varying phase estimation technology \cite{xu2024robust, liao2025high} enables polychromatic Local Local Oscillator (LLO) coherent reception, removing the requirement for time-frequency locking networks \cite{droste2013optical, chen2024dual}. Second is network capacity. The polychromatic distribution scheme, combined with the state broadcasting method, expands the network capacity to hundreds of users. Each user can perform beacon-aided LLO coherent reception and obtain independent final keys after reconciliation and privacy amplification. Last is SKR. The Bose-Einstein distribution of the thermal state provides high-bandwidth randomness, enabling ultra-high repetition rates. Polychromatic distribution mitigates the SKR loss caused by state broadcasting, allowing hundreds of users to perform $\mathrm{Mbps}$-level information encryption. In experiments, we constructed a TS-QAN with a capacity of $304$. Under the covariance-matrix-based network security calculation including Holevo leakage and residual broadcast-induced correlations \cite{xu2026polychromatic}, the experimental results are: the SKR of $13.76$ $\mathrm{Gbps}$ over $5$ $\mathrm{km}$, $7.28$ $\mathrm{Gbps}$ over $15$ $\mathrm{km}$, and $2.47$ $\mathrm{Gbps}$ over $30$ $\mathrm{km}$ in the asymptotic case; the SKR of $3.60$ $\mathrm{Gbps}$ over $5$ $\mathrm{km}$ and $0.26$ $\mathrm{Gbps}$ over $15$ $\mathrm{km}$ under finite-size effect. They validate the feasibility of TS-QAN, satisfying the performance requirements of the advanced and dominant CAN.

In this paper, we begin by describing the fundamental architecture of the TS-QAN. Then, it is verified from both theoretical and experimental perspectives. The theoretical aspect comprises the Prepare-Measure (PM) model and the Entanglement-Based (EB) model. The experimental aspect encompasses the experimental setup and the experimental results. Finally, we present the discussion and conclusion.

\section*{Results}

\subsection*{Thermal State Quantum Key Distribution Access Network}

Based on quantum information theory, QAN can provide secure encryption for CAN through OTP. OTP requires QAN and CAN to operate at the same information rate and network capacity. However, existing QANs struggle to meet this condition. Let's explore this challenging situation by beginning with an introduction to the existing QANs.

Similar to OLT, ODN, and ONU in CAN, QAN also has Quantum Line Terminal (QLT), Quantum Distribution Network (QDN), and Quantum Network Unit (QNU). Since QKD can be divided into Discrete-Variable QKD (DV-QKD) and Continuous-Variable QKD (CV-QKD) based on the encoding Hilbert space, QAN can also be divided into Discrete-Variable QAN (DV-QAN) and Continuous-Variable QAN (CV-QAN) in the same way. DV-QAN \cite{townsend1997quantum, choi2011quantum, frohlich2013quantum, frohlich2015quantum, wang2021practical, huang2024cost}, encoded in a finite-dimensional Hilbert space, typically approximates coherent states as the single photon states at the QLT (security is ensured by introducing decoy states \cite{wang2005beating, peng2007experimental}). Then, the random numbers generated by QRNGs are modulated on single photon states via phase encoding. Next, these single photon states are distributed through the multiplexer of the QDN. Last, each QNU measures them using a single photon detector. CV-QAN \cite{huang2021realizing, wang2023experimental, xu2023round, hajomer2024continuous, li2024experimental, qi2024experimental, pan2025high, xu2026polychromatic}, encoded in an infinite-dimensional Hilbert space, typically attenuates coherent states into weak coherent states at the QLT \cite{grosshans2002continuous}. Then, the random numbers generated by QRNGs are modulated on the weak coherent states via Gaussian encoding. Next, these weak coherent states are distributed through the multiplexer of the QDN. Last, each QNU measures them using a coherent detector. Therefore, both DV-QAN and CV-QAN can be categorized as coherent state networks, as shown in the upper part of Fig. \ref{fig1a}. Coherent state networks utilize coherent states as the light source, achieving point-to-multipoint QKD through modulators and multiplexers. The SKR per user of coherent state networks is at the $\mathrm{Mbps}$ level, capable of real-time encryption for voice and messages. The maximum number of users that the network can accommodate is about several tens. The reliance on active modulation brings about the need for modulators and QRNGs. The data used here, $765.14$ $\mathrm{kbps}$ over $5$ $\mathrm{km}$, is given by the state-of-the-art BS-QAN based on state broadcasting method with a network capacity of $32$ and a repetition rate of $1$ $\mathrm{GHz}$. The performance of the coherent state network ($24.48\mathrm{-Mbps}$ $@$ $32\mathrm{-users}$) cannot meet the advanced and dominant CAN's demand for communication rate and network capacity ($10\mathrm{-Gbps}$ $@$ $256\mathrm{-users}$).

\begin{figure*}[!h]
	\centering
	\includegraphics[width=1\linewidth]{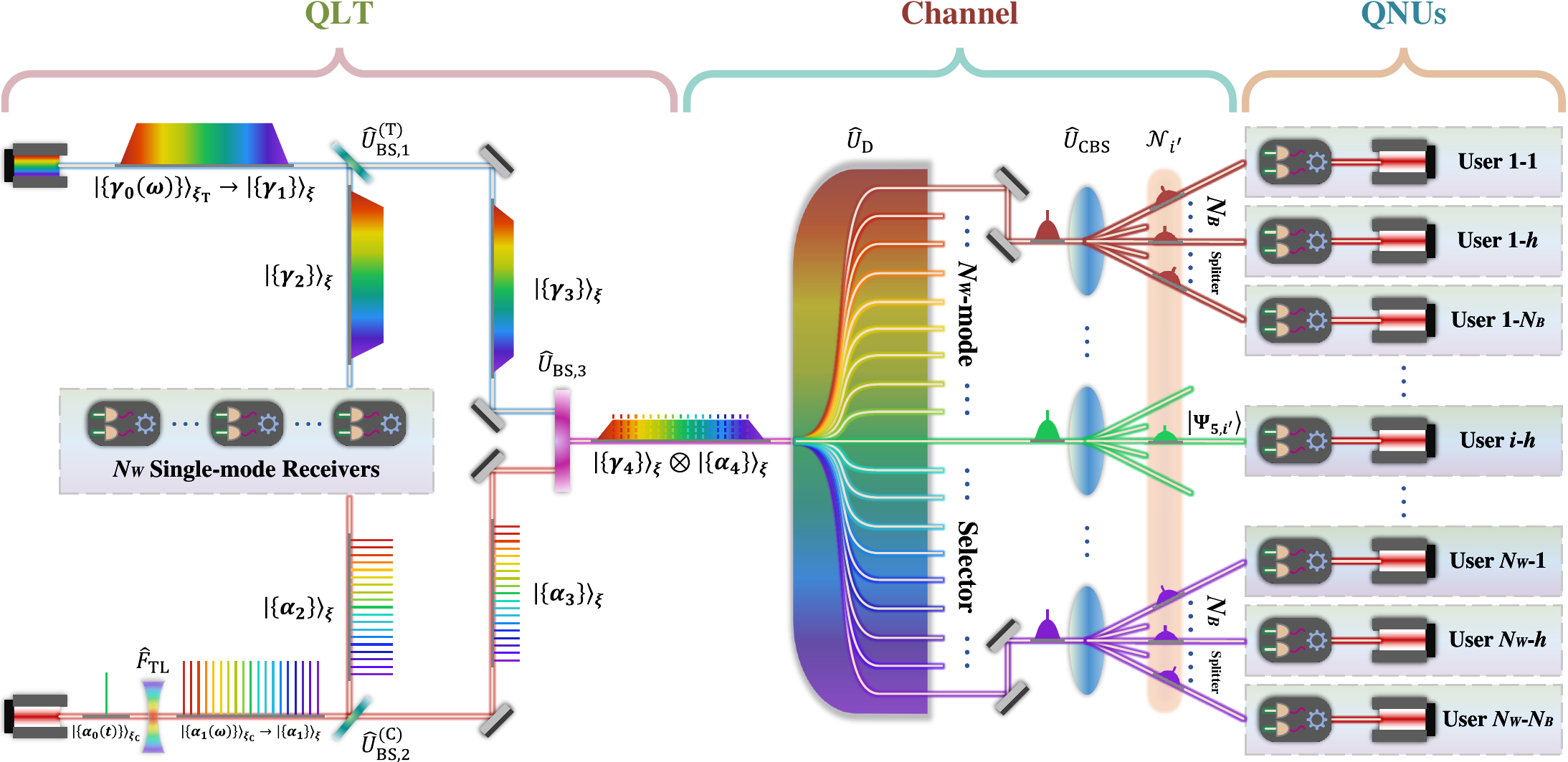}
	\caption{PM model of TS-QAN. It provides the correspondence between indices, where $i \in \left\{1,2,\cdots, N_W \right\}$ represents the index of the frequency modes, $h \in \left\{1,2,\cdots, N_B \right\}$ represents the index of the power-splitting branches, and $i^{\prime} \in \left\{1,2,\cdots, N_W \cdot N_B \right\}$ represents the index of the users. The $i^{\prime}$-th user is referred to as User $i-h$, where $i=\left\lfloor \left(i^{\prime}-1\right)/N_B \right\rfloor+1$, $h=\left(i^{\prime}-1\right) \bmod N_B+1$, and equivalently $i^{\prime}=\left(i-1\right)N_B+h$. The maximum number of users, that is, the network capacity, is $N=N_W \cdot N_B$. In the schematic, the $N_W$-mode selector denotes the passive frequency-allocation operation $\hat U_D$, whose mode-domain matrix representation is $D_{\rm full}$. The $N_B$-splitter denotes the physical cascaded beam-splitting operation $\hat U_{\rm CBS}$, whose port-space matrix representation is $U_{\rm CBS}$. The residual user-channel map is denoted by $\mathcal N_{i'}(T_{d,i'},\varepsilon_{i'},\theta_{p,i'})$, including residual transmittance, Alice-referred excess noise, and phase rotation.}
	\label{fig2}
\end{figure*}

The TS-QAN proposed in this article addresses the long-standing issue with QANs. The polychromatic thermal states for TS-QAN are provided by a broadband thermal source at QLT. Then, the QLT's key is acquired via passive preparation structures. Next, these polychromatic thermal states are distributed through the multiplexer of the QDN. Last, each QNU obtains the QNU's key using a coherent detector. The thermal state network primarily utilizes two characteristics of the thermal source to achieve high performance in point-to-multipoint QKD. The first is the abundant polychromatic resources. The polychromatic resources of the thermal state network stem from the broad spectrum of the thermal source, whereas the coherent state network requires complex polychromatic resource preparation techniques. The second is the high-bandwidth randomness. The randomness of the thermal state network stems from the Bose-Einstein distribution of the thermal source, while the coherent state network requires high-performance modulators and QRNGs. The brief steps and main features of the thermal state network are shown in the lower part of Fig. \ref{fig1a}. The SKR per user of thermal state network is at the $10\mathrm{Mbps}$ level, capable of real-time encryption for videos and live-streams. The maximum number of users that the network can accommodate is about several hundred. The reliance on passive preparation eliminates the need for modulators and QRNGs, and the number of passive preparation structures is much less than the network capacity. The data used here, $45.26$ $\mathrm{Mbps}$ over $5$ $\mathrm{km}$, is given by the experimental results of TS-QAN with a network capacity of $304$. The performance of the thermal state network ($13\mathrm{-Gbps}$ $@$ $304\mathrm{-users}$) fully meets the advanced and dominant CAN's demands for communication rate and network capacity ($10\mathrm{-Gbps}$ $@$ $256\mathrm{-users}$).

In the following, we will introduce the general implementation and topology of the thermal state network, as illustrated in Fig. \ref{fig1b}. The thermal state network contains two essential components: a thermal source and a flat optical comb. In this architecture, the broadband thermal source provides the distributed quantum-randomness resource, whereas the flat optical comb serves as a polychromatic beacon for coherent reception. The polychromatic thermal states generated by the thermal source provide sufficient polychromatic quantum randomness for key generation among $304$ $\mathrm{users}$. The polychromatic beacons produced by the flat optical comb give $304$ $\mathrm{users}$ the ability to extract keys. Both are indispensable. The distribution strategy of the thermal state network can be understood based on the division of the thermal source. The spectrum of polychromatic thermal states is equally divided into $19$ blocks horizontally, corresponding to $19$ frequency modes, and is equally split into $16$ blocks vertically, corresponding to $16$ power-splitting branches. The number $a$ on the spectrum indicates that the photons from this block are allocated to the $a$-th user, where $a \in \left\{1,2,\cdots,304\right\}$. QLT, equipped with a thermal source and a flat optical comb, is responsible for passive preparation and polychromatic distribution to evenly divide the thermal states into $19$ frequency modes. Their waveforms are denoted as the large wave packets of different colors in the figure. QDN adopts a physical power splitter to evenly split the thermal state of each frequency mode into $16$ power-splitting branches. Their waveforms are depicted as the small wave packets of the same color in the diagram. QNUs employ coherent receivers to detect these $304$ thermal states. They are decoded into mutually independent final keys through the state broadcasting method, with residual broadcast-induced correlations accounted for in the SKR calculation. The topology of the thermal state network can be implemented in various forms. User nodes can be arranged around the center node to form a star topology. Alternatively, user nodes can be concentrated on one side of the center node to form a tree topology. Furthermore, depending on the coverage requirements, the positions of the center node, distribution nodes, and user nodes can be freely adjusted to create a network topology suitable for the practical scenario.

\subsection*{Prepare-Measure Model}

Continuous-mode quantum optics \cite{raymer1989temporal, blow1990continuum, fabre2020modes, raymer2020temporal, chen2023continuous, sun2025analyzing} allows for clear differentiation of the numerous quantum states in quantum networks according to their characteristics. In the previous work \cite{xu2026polychromatic}, we have developed a multimode quantum network theory based on continuous-mode quantum optics, including universal network evolution operators and typical network evolution processes. From preparation to measurement, the PM model describing the evolution process of TS-QAN will be presented within the framework of the multimode quantum network theory. As depicted in Fig. \ref{fig2}, the PM model can be divided into three parts: QLT, channel, and QNUs.

\emph{\textbf{QLT}} ----- The spontaneous emission source at QLT emits a multimode thermal state in its native thermal-source wave-packet basis $\bm{\xi}_{\mathrm{T}}=\{\xi_{\mathrm{T},i}\}_{i=1}^{N_W}$. Since a thermal state is a mixed state, its continuous-mode density operator is first written in the Fock basis as
\begin{equation}
	\label{eq1}
	\hat{\rho}_{\mathrm{T}}
	=
	\bigotimes_{i=1}^{N_W}
	\sum_{n=0}^{\infty}
	P_i\left(n\right)
	\left|n\right\rangle_{\xi_{\mathrm{T},i}}
	\left\langle n\right|_{\xi_{\mathrm{T},i}} .
\end{equation}
Here, $\xi_{\mathrm{T},i}$ denotes the thermal-source wave packet of the $i$-th frequency mode, $\bar n_i$ is the mean photon number of this mode, and $P_i\left(n\right)=\bar{n}_i^{\,n}/\left(1+\bar{n}_i\right)^{n+1}$ is the Bose-Einstein photon-number distribution. According to the Weisskopf-Wigner theory \cite{scully1997quantum}, the spectral envelope can be written as $\xi^{0}_{\mathrm{T},i}\left(\omega\right)=g^*_i/\left(\delta_{\omega}+\jj\Gamma/2\right)$, where $g_i$ is the coupling coefficient of the $i$-th mode, $\delta_{\omega}$ is the detuning between the atom and the radiation field, and $\Gamma$ is the dissipation parameter. Referencing the Glauber--Sudarshan $P$ representation \cite{scully1997quantum}, the same multimode thermal state can be represented as a Gaussian ensemble of coherent-state realizations,
\begin{equation}
	\scalebox{0.89}
	{$
	\hat{\rho}_{\mathrm{T}}
	=
	\int
	\left[
	\prod_{i=1}^{N_W}
	\frac{1}{\pi V_{0,i}}
	\exp\!\left(-\frac{|\gamma_{0,i}|^2}{V_{0,i}}\right)
	\mathrm{d}^{2}\gamma_{0,i}
	\right]
	\bigotimes_{i=1}^{N_W}
	\left|\gamma_{0,i}\right\rangle_{\xi_{\mathrm{T},i}}
	\left\langle\gamma_{0,i}\right|_{\xi_{\mathrm{T},i}} .
	$}
\end{equation}
This equation means that each experimental realization of the multimode thermal field is represented by a tensor product of coherent states sampled from the Gaussian distribution of $\{\gamma_{0,i}\}$, which is compactly denoted in Fig.~\ref{fig2} as $\left|\left\{\gamma_0(\omega)\right\}\right\rangle_{\bm{\xi}_{\mathrm{T}}}$. Here, $\gamma_{0,i}=x_{0,i}+\jj p_{0,i}$ and $V_{0,i}=\left\langle |\gamma_{0,i}|^2\right\rangle$. The full thermal state is recovered by ensemble averaging. The native thermal wave packet can be written as $\xi_{\mathrm{T},i}\left(\omega\right)=\xi^0_{\mathrm{T},i}\left(\omega\right)\delta\left(\omega-\omega_{\mathrm{T},i}\right)$, where $\omega_{\mathrm{T},i}$ denotes the centre frequency of the $i$-th thermal-source frequency mode. The repetition rate is $F_r=\Delta\omega/2\pi$.

Meanwhile, the single-frequency-mode laser at QLT emits a multi-temporal-mode coherent beacon state in its native temporal coherent-state basis $\bm{\xi}_{\mathrm{C}}^{(t)}=\{\xi_{\mathrm{C},m}^{(t)}\}$, which can be written as $\left|\left\{\alpha_0(t)\right\}\right\rangle_{\bm{\xi}_{\mathrm{C}}^{(t)}}=\bigotimes_m\left|\alpha_{0,m}\right\rangle_{\xi_{\mathrm{C},m}^{(t)}}$. It is converted into a multi-frequency-mode coherent beacon state $\left|\left\{\alpha_1(\omega)\right\}\right\rangle_{\bm{\xi}_{\mathrm{C}}^{(\omega)}}=\bigotimes_{i=1}^{N_W}\left|\alpha_{1,i}\right\rangle_{\xi_{\mathrm{C},i}^{(\omega)}}$ by the time-lens Fourier-transform operator $\hat F_{\mathrm{TL}}\!\left(t=\omega/\ddot{\phi}_t\right)$, consistent with the definition in the Supplementary Information. Taking the frequency-domain beacon wave-packet basis $\bm{\xi}_{\mathrm{C}}^{(\omega)}$ as the mode reference, we re-orthonormalise the thermal-source basis $\bm{\xi}_{\mathrm{T}}$ and use the common frequency-wave-packet basis $\bm{\xi}=\{\xi_i\}_{i=1}^{N_W}$ in the subsequent PM and EB descriptions. The thermal-source realization $\left|\left\{\gamma_0(\omega)\right\}\right\rangle_{\bm{\xi}_{\mathrm{T}}}$ is thereby transformed and relabelled as $\left|\left\{\gamma_1\right\}\right\rangle_{\bm{\xi}}$, with amplitudes $\gamma_{1,i}=x_i+\jj p_i$. The common wave packet may be written as $\xi_i\left(\omega\right)=\xi_i^0\left(\omega\right)\delta\left(\omega-\omega_i\right)$, where $\omega_i$ is the centre frequency of the $i$-th common frequency mode. In this common basis, $\left|\left\{\gamma_1\right\}\right\rangle_{\bm{\xi}}=\bigotimes_{i=1}^{N_W}\left|\gamma_{1,i}\right\rangle_{\xi_i}$ and $\left|\left\{\alpha_1\right\}\right\rangle_{\bm{\xi}}=\bigotimes_{i=1}^{N_W}\left|\alpha_{1,i}\right\rangle_{\xi_i}$.
The multi-frequency-mode thermal-state realization $\left|\left\{\gamma_1\right\}\right\rangle_{\bm{\xi}}$ and coherent beacon state $\left|\left\{\alpha_1\right\}\right\rangle_{\bm{\xi}}$ are each split into two beams by respective unbalanced beam-splitter operations $\hat U_{{\rm BS},1}^{({\rm T})}$ and $\hat U_{{\rm BS},2}^{({\rm C})}$. The high-power beams of both, $\left|\left\{\gamma_2\right\}\right\rangle_{\bm{\xi}}$ and $\left|\left\{\alpha_2\right\}\right\rangle_{\bm{\xi}}$, are input into $N_W$ single-mode receivers for detection. The single-mode thermal-state realization $\left|\gamma_{2,i}\right\rangle_{\xi_i}$ and coherent beacon state $\left|\alpha_{2,i}\right\rangle_{\xi_i}$ are regarded as the original signal and the QLT's LO, respectively, with the measured passive-preparation variance $V_{\mathrm{P},i}=\left\langle |\gamma_{2,i}|^2\right\rangle$. The second moment $\sigma^2_{\mathrm{A},i}$ of the $i$-th reception result $\hat{D}_{\mathrm{A},i}$ is given by
\begin{equation}
	\label{eq3}
	\begin{aligned}
		\sigma^2_{\mathrm{A},i}
		=\left\langle \hat{D}_{\mathrm{A},i}\hat{D}^{\dagger}_{\mathrm{A},i}\right\rangle
		=\eta_e\eta_{\mathrm{A},i} V_{\mathrm{P},i}+{v_{\rm el}}+1 .
	\end{aligned}
\end{equation}
Here, $\eta_e$ denotes the quantum detection efficiency, ${v_{\rm el}}$ represents the variance of electronic noise, and $\eta_{\mathrm{A},i}$ is the effective mode-matching factor of the QLT receiver. In the ideal calibrated single-mode condition used for the experimental reconstruction, $\eta_{\mathrm{A},i}$ is absorbed into the measured value of $V_{\mathrm{P},i}$ or set to unity. The low-power beams of the multimode thermal-state realization $\left|\left\{\gamma_3\right\}\right\rangle_{\bm{\xi}}$ and coherent beacon state $\left|\left\{\alpha_3\right\}\right\rangle_{\bm{\xi}}$ are input into the balanced beam-splitter operation $\hat U_{{\rm BS},3}$ for coupling, forming a tensor product state $\left|\left\{\gamma_4\right\}\right\rangle_{\bm{\xi}} \otimes \left|\left\{\alpha_4\right\}\right\rangle_{\bm{\xi}}$. At this point, the equivalent modulation variance of passive preparation is mode dependent and is estimated as the ensemble variance $V_{\mathrm{A},i}=\left\langle |\gamma_{4,i}|^2\right\rangle$. In an ideal symmetric regime, these values can be approximately equalized across channels, but they are not assumed to be mathematically identical for every wavelength channel. Subsequently, the tensor product state is transmitted into the channel.

\begin{figure*}[!h]
	\centering
	\includegraphics[width=1\linewidth]{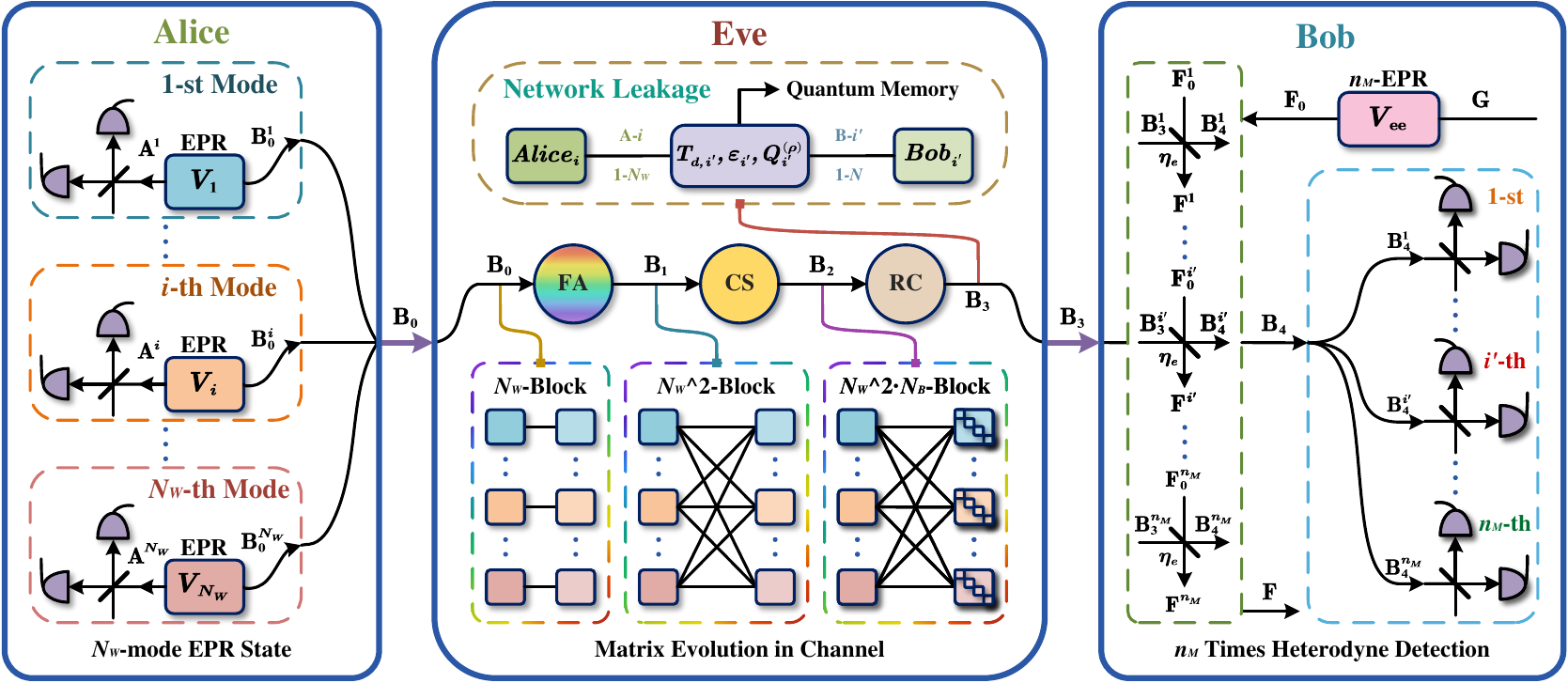}
	\caption{EB model of TS-QAN. In this schematic only, FA denotes the frequency-allocation stage, whose strict matrix representation is $D_{\rm full}$; CS denotes the cascaded-splitting stage, whose strict port-space matrix representation is $U_{\rm CBS}$; and RC denotes the residual user channel. The labels $N_W^2$-Block and $N_W^2N_B$-Block denote Alice--Bob covariance blocks after the frequency-allocation and cascaded-splitting stages, respectively, rather than newly generated EPR pairs. The $n_M$-EPR block and $n_M$ times heterodyne detection represent the trusted heterodyne detection model for the measured block $M$, where $n_M$ depends on the selected receiver-security policy. Alice corresponds to QLT, Eve corresponds to the channel/security leakage, and Bob corresponds to QNUs.}
	\label{fig3}
\end{figure*}

\emph{\textbf{Channel}} ----- The tensor product state first enters the passive frequency-allocation operation $\hat U_D$, whose mode-domain matrix representation is the real orthogonal matrix $D_{\rm full}$. We write $D_{ku}\equiv[D_{\rm full}]_{ku}$ for compactness. The $k$-th frequency output can therefore contain coherent contributions from all source-side frequency modes. This stage is followed by the physical cascaded beam-splitting operation $\hat U_{\rm CBS}$, whose port-space matrix representation is $U_{\rm CBS}$ and whose branch ratio is $r_h$, and finally by the residual user-channel map $\mathcal N_{i'}(T_{d,i'},\varepsilon_{i'},\theta_{p,i'})$. For the $i^{\prime}$-th user, we use $k=k\left(i^{\prime}\right)=\left\lfloor \left(i^{\prime}-1\right)/N_B \right\rfloor+1$ and $h=h\left(i^{\prime}\right)=\left(i^{\prime}-1\right)\bmod N_B+1$. In the PM amplitude, the deterministic part of $\mathcal N_{i'}$ contributes the factor $e^{-\jj\theta_{p,i'}}\sqrt{T_{d,i'}}$, whereas the excess-noise contribution is retained in the covariance-matrix description. Considering the residual attenuation and phase rotation, the component-resolved signal and beacon fields received by User $i^{\prime}$ are written as
\begin{equation}
	\scalebox{1}
	{$
	\begin{aligned}
		\left|\Psi_{5,i^{\prime}}\right\rangle
		&=
		\bigotimes_{u=1}^{N_W}
		\left|\gamma_{5,i^{\prime}\leftarrow u}\right\rangle_{\xi_u}^{(i^{\prime},\mathrm{s})}
		\otimes
		\left|\alpha_{5,i^{\prime}\leftarrow u}\right\rangle_{\xi_u}^{(i^{\prime},\mathrm{b})},\\
		\gamma_{5,i^{\prime}\leftarrow u}
		&=e^{-\jj\theta_{p,i^{\prime}}}\sqrt{T_{d,i^{\prime}} r_h}\,D_{ku}\gamma_{4,u},\\
		\alpha_{5,i^{\prime}\leftarrow u}
		&=e^{-\jj\theta_{p,i^{\prime}}}\sqrt{T_{d,i^{\prime}} r_h}\,D_{ku}\alpha_{4,u} .
	\end{aligned}
	$}
\end{equation}
Thus, finite frequency isolation is not represented by a single output wave packet only; instead, the user input contains a set of frequency components $\{\xi_u\}$ whose weights are determined by $D_{ku}$. After the channel, $N_WN_B$ user ports are sent to the QNUs.

\emph{\textbf{QNUs}} ----- The $i^{\prime}$-th QNU prepares a laser that can emit a single-frequency-mode coherent state $\left|\alpha_{{\rm LO},i^{\prime}}\right\rangle_{\xi_{\rm LO}^{(i^{\prime})}}$. The received multimode state $\left|\Psi_{5,i^{\prime}}\right\rangle$ and the LO define an effective coherent-detection mode through the receiver response, shot-noise normalization, and DSP. The second moment of the reception result $\hat{D}_{\mathrm{B},i^{\prime}}$ is therefore written as
\begin{equation}
	\label{eq5}
	\begin{aligned}	
		\sigma^2_{\mathrm{B},i^{\prime}}
		&=\left\langle\Delta\hat{D}_{\mathrm{B},i^{\prime}}\Delta\hat{D}_{\mathrm{B},i^{\prime}}^{\dagger}\right\rangle \\
		&=\eta_e\sum_{u,v=1}^{N_W}
		m_{i^{\prime}\leftarrow u}m^*_{i^{\prime}\leftarrow v}
		\left[\Gamma_{\mathrm{B},i^{\prime}}^{(\mathrm{in})}\right]_{uv}
		+{v_{\rm el}}+1 .
	\end{aligned}
\end{equation}
Here, $m_{i^{\prime}\leftarrow u}=\int \mathrm{d}\omega\,[\xi_{\mathrm{rec}}^{(i^{\prime})}(\omega)]^*\xi_u(\omega)$ is the overlap between the DSP-defined receiver mode and the $u$-th frequency component. The ideal local single-frequency receiver is recovered when the receiver mode selects only the target component. For a global receiver that resolves several frequency components, the corresponding multimode second-moment matrix is represented by $\Sigma_{\mathrm{B},i^{\prime}}^{(\mathrm{glob})}=\eta_eM_{\mathrm{mm},i^{\prime}}\Gamma_{\mathrm{B},i^{\prime}}^{(\mathrm{in})}M_{\mathrm{mm},i^{\prime}}^{\dagger}+\left(v_{\rm el}+1\right)I$. The DSP kernel can be written as $f_{\mathrm{DSP}}^{(i^{\prime})}(t)=\exp\!\left\{\jj\left[\omega_{\mathrm{LO},i^{\prime}}-\omega_k\right]t-\jj\theta_{p,i^{\prime}}(t)\right\}$ and is used together with the shot-noise normalization and receiver response to define the effective receiver wave packet. For the full construction of the DSP-defined receiver mode, please refer to the Supplementary Information and Refs. \cite{xu2024robust, liao2025high}.

The next step is data processing of TS-QAN, which involves parameter estimation, reconciliation decoding, and privacy amplification. The QLT quadrature data are labeled as $A_i=\left(X_{\mathrm{A},i},P_{\mathrm{A},i}\right)$. For a local receiver, the $i^{\prime}$-th QNU quadrature data are labeled as $B_{i^{\prime}}^{(\mathrm{loc})}=\left(X_{\mathrm{B},i^{\prime}},P_{\mathrm{B},i^{\prime}}\right)$; for a global receiver, the retained Bob-side data form a multimode vector $\mathbf{B}_{i^{\prime}}^{(\mathrm{glob})}=\left(B_{i^{\prime}\leftarrow1},B_{i^{\prime}\leftarrow2},\ldots,B_{i^{\prime}\leftarrow N_W}\right)$. During parameter estimation, the passive-preparation data are used to estimate the equivalent modulation variance $V_{\mathrm{A},i}=\left\langle |\gamma_{4,i}|^2\right\rangle$, and the excess-noise parameters are estimated from the calibrated user data \cite{qi2018passive, qi2020experimental}. In the reconciliation decoding \cite{leverrier2008multidimensional}, since the data $A=\left\{A_1,A_2,\cdots,A_{N_W}\right\}$ of QLT and the data $B=\left\{B_1,B_2,\cdots,B_{N_W \cdot N_B}\right\}$ of the QNUs are not symmetric in terms of quantity, the QLT replicates its data $N_B$ times to form the user-indexed data $\widetilde A=\left\{\widetilde A_1,\widetilde A_2,\cdots,\widetilde A_{N_W \cdot N_B}\right\}$, which becomes quantitatively symmetric with the QNUs' data. Each QNU calculates the mapping matrix and syndrome according to independent random numbers, the check matrix, and the data $B$. QNUs send the mapping matrices and syndromes to the QLT via classical channels. The random numbers and mapping matrices are mutually independent, enabling state broadcasting \cite{hajomer2024continuous, pan2025high} without direct reuse of identical reconciliation mappings; residual inter-user classical correlations are accounted for by $I_{\mathrm{res},i^{\prime}}$ in the SKR formula. QLT decodes using mapping matrices, syndromes, and the replicated data $\widetilde A$. Frames that are correctly decoded are retained. Finally, privacy amplification for correct frames is performed through hash compression, generating $N_W \cdot N_B$ pairs of independent secure keys.

\subsection*{Entanglement-Based Model}

Multimode density matrix provides a rigorous performance assessment by considering the quantum network as a whole. In the previous work \cite{xu2026polychromatic}, a general EB model based on the multimode density matrix has been presented, thereby providing schemes for solving network SKR under different regimes. Since TS-QAN employs the state broadcasting method \cite{hajomer2024continuous, pan2025high}, the EB model also needs to consider the leakage of classical information among users. As illustrated in Fig. \ref{fig3}, the EB model can be divided into three parts: Alice, Eve, and Bob.

\emph{\textbf{Alice}} ----- First of all, Alice prepares the $N_W$-mode Einstein-Podolsky-Rosen (EPR) state $\mathrm{A}\mathrm{B}_0$, with a variance $V_i=V_{\mathrm{A},i}+1$ of the $i$-th mode. Each single-mode part $\mathrm{A}^i$ is retained and detected by Alice, while each single-mode part $\mathrm{B}_0^i$ is sent to Bob.

\emph{\textbf{Eve}} ----- In the channel controlled by Eve, there are three evolution stages: frequency allocation, cascaded splitting, and residual channel, which are consistent with the PM model and the Supplementary Information. For compactness, Fig. \ref{fig3} labels the successive Bob-side blocks as $B_0,B_1,B_2$, and $B_3$; these schematic labels correspond respectively to the source-side Bob block, the frequency-allocated block, the cascaded-splitting output block, and the residual-channel output block used in the covariance-matrix formulation. Under different evolution stages, the entanglement connections of the EPR state change. When initially entering the channel, the EPR state $\mathrm{A}\mathrm{B}_0$ consists of $N_W$ EPR pairs. After the frequency-allocation stage, the Bob-side outputs remain $N_W$ frequency-allocated ports, while the Alice--Bob correlation block becomes dense. After the physical power-splitting stage, each frequency output is further divided into $N_B$ power branches, giving $N=N_WN_B$ Bob user modes and $N_W^2N_B$ effective Alice--Bob covariance blocks rather than newly generated EPR pairs. Then, Eve performs a collective attack on the results $\mathrm{A}\mathrm{B}_3$ output from the residual channel. Although the multi-dimensional density matrix of $\mathrm{A}\mathrm{B}_3$ cannot be directly displayed, it can be fully expressed by the following quantum covariance matrices. Here, $u\in\{1,2,\ldots,N_W\}$ denotes the Alice-side source frequency-mode index, $i^{\prime}\in\{1,2,\ldots,N_WN_B\}$ denotes the Bob-side user index, and $i^{\prime}=(k-1)N_B+h$ denotes User $k-h$. The EPR variance of the $u$-th frequency mode is $V_u=V_{\mathrm{A},u}+1$, $C_u=\sqrt{V_u^2-1}$, $I_2=\mathrm{diag}(1,1)$, and $Z=\mathrm{diag}(1,-1)$. The passive frequency-allocation matrix is $D_{\rm full}$ with elements $D_{ku}\equiv[D_{\rm full}]_{ku}$, and the physical splitting ratio of branch $h$ is $r_h$. The Alice--Bob covariance block after the full channel is
\begin{equation}
	\scalebox{1}
	{$
	\begin{aligned}
		\Psi_{\mathrm{A}_u\mathrm{B}_{i^{\prime}}}
		&=
		\begin{bmatrix}
			V_u I_2 & V^{\mathrm{net}}_{\mathrm{A}_u\mathrm{B}_{i^{\prime}}}\\
			\left(V^{\mathrm{net}}_{\mathrm{A}_u\mathrm{B}_{i^{\prime}}}\right)^{T} & V^{\mathrm{net}}_{\mathrm{B}_{i^{\prime}}\mathrm{B}_{i^{\prime}}}
		\end{bmatrix},\\
		V^{\mathrm{net}}_{\mathrm{A}_u\mathrm{B}_{i^{\prime}}}
		&=\sqrt{T_{d,i^{\prime}}r_h}\,C_uD_{ku}Z .
	\end{aligned}
	$}
\end{equation}
The Bob-side covariance blocks are obtained by first applying the physical splitter and then applying the residual user channels,
\begin{equation}
	\scalebox{0.89}
	{$
	\begin{aligned}
		V^{\mathrm{CBS}}_{\mathrm{B}_k^{(h)}\mathrm{B}_l^{(g)}}
		&=
		\left[
		\sqrt{r_hr_g}\sum_{u=1}^{N_W}D_{ku}D_{lu}V_u
		+\delta_{kl}\left(\delta_{hg}-\sqrt{r_hr_g}\right)
		\right]I_2,\\
		V^{\mathrm{net}}_{\mathrm{B}_{i^{\prime}}\mathrm{B}_{\ell^{\prime}}}
		&=\sqrt{T_{d,i^{\prime}}T_{d,\ell^{\prime}}}
		V^{\mathrm{CBS}}_{\mathrm{B}_k^{(h)}\mathrm{B}_l^{(g)}}
		+\delta_{i^{\prime}\ell^{\prime}}
		\left(1-T_{d,i^{\prime}}+T_{d,i^{\prime}}Q_{i^{\prime}}^{(\rho)}{\varepsilon_{i^{\prime}}}\right)I_2,
	\end{aligned}
	$}
\end{equation}
where $\ell^{\prime}=(l-1)N_B+g$, ${\varepsilon_{i^{\prime}}}$ is the Alice-referred excess noise, and $\rho\in\{\mathrm{loc},\mathrm{glob}\}$ denotes the receiver capability. The receiver-dependent signal-power factor is $Q_{i^{\prime}}^{(\rho)}=r_hq_i^{(\rho)}$ with $i=k(i^{\prime})$, where $q_i^{(\mathrm{loc})}=|D_{ii}|^2$ and $q_i^{(\mathrm{glob})}=1$. The two-user Bob covariance matrix can then be written compactly as
\begin{equation}
	\begin{aligned}
		\Psi_{\mathrm{B}_{i^{\prime}}\mathrm{B}_{\ell^{\prime}}}
		=
		\begin{bmatrix}
			V^{\mathrm{net}}_{\mathrm{B}_{i^{\prime}}\mathrm{B}_{i^{\prime}}} & V^{\mathrm{net}}_{\mathrm{B}_{i^{\prime}}\mathrm{B}_{\ell^{\prime}}}\\
			V^{\mathrm{net}}_{\mathrm{B}_{\ell^{\prime}}\mathrm{B}_{i^{\prime}}} & V^{\mathrm{net}}_{\mathrm{B}_{\ell^{\prime}}\mathrm{B}_{\ell^{\prime}}}
		\end{bmatrix} .
	\end{aligned}
\end{equation}
The off-diagonal Bob--Bob blocks are generally nonzero because the frequency-allocation matrix and the physical splitter create shared covariance blocks across user modes. The residual-channel output block $\mathrm{B}_3$ in the schematic is then passed to the Bob-side receiver model.

\begin{figure*}[!h]
	\centering
	\includegraphics[width=1\linewidth]{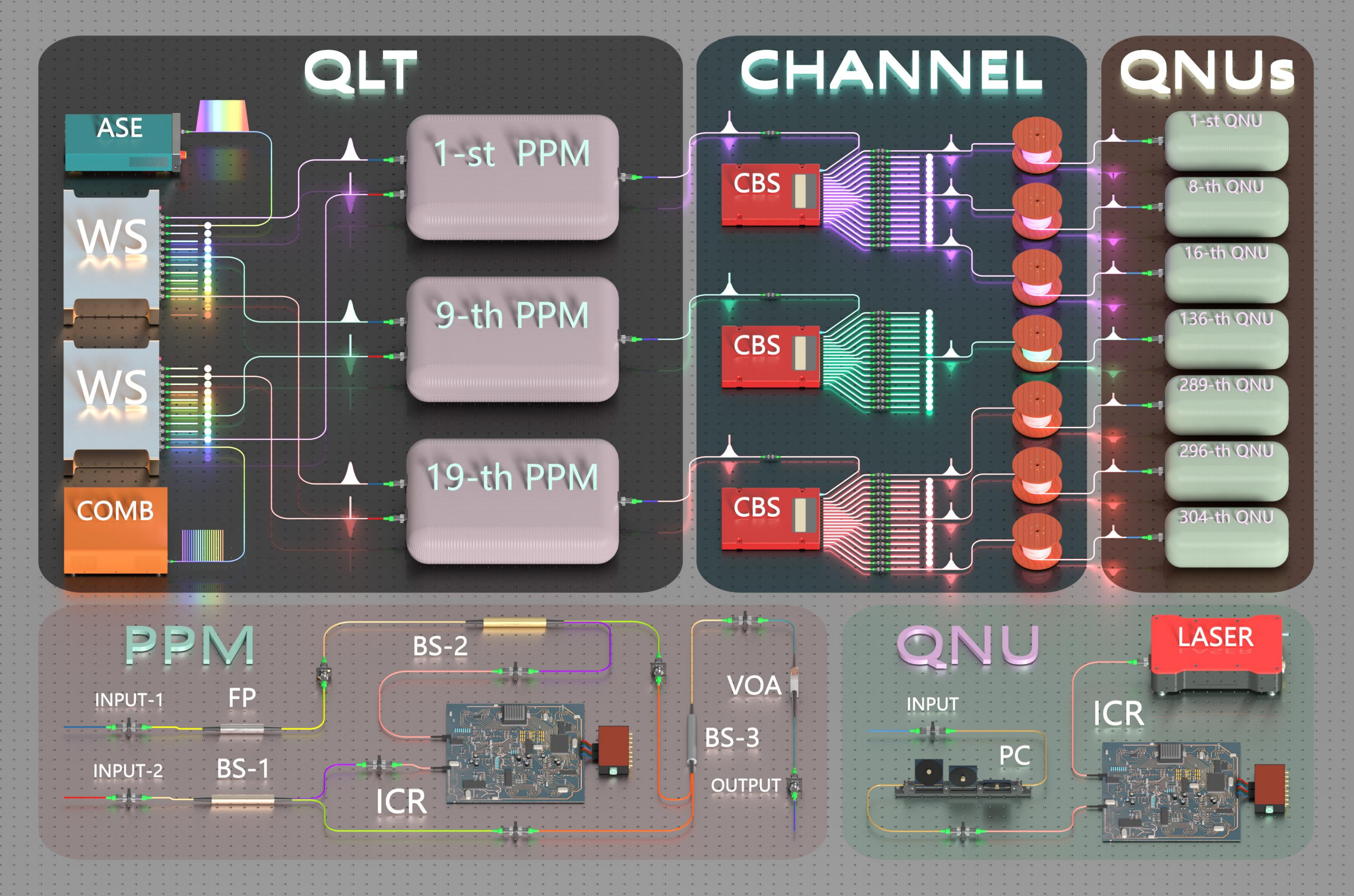}
	\caption{Experimental setup of TS-QAN. In the WS, the gradient-colored fiber represents input, and the solid-colored fibers denote outputs. $\mathrm{BS-}1$ and $\mathrm{BS-}2$ are $10:90$ $1 \times 2$ BS and $1:99$ $1 \times 2$ BS, respectively, each equipped with a yellow input fiber, a purple high-power output fiber, and a green low-power output fiber. $\mathrm{BS-}3$ is a $50:50$ $2 \times 1$ BS, featuring two orange input fibers and one yellow output fiber.}
	\label{fig4}
\end{figure*}

\emph{\textbf{Bob}} ----- Bob receives and measures the selected measured block $M$. Depending on the receiver-security policy, the non-target Bob modes are either assigned to Eve, retained as trusted but unmeasured systems, or included in the measured Bob-side set. Thus, the trusted-user setting is one security partition rather than a default assumption for all calculations. The auxiliary system $\mathrm{F}_0\mathrm{G}$ comprises $n_M$ EPR pairs, where $n_M=1$ for a local single-user receiver, $n_M=N_W$ for an ideal global single-user receiver, $n_M=N$ for the local all-measured policy, and $n_M=NN_W$ for the fully frequency-resolved global all-measured policy. If a physical multimode receiver implements an equivalent unitary compression, the same policy can be represented in the corresponding reduced but symplectically equivalent basis. Its covariance matrix is
\begin{equation}
	\begin{aligned}
		V_{\mathrm{F}_0\mathrm{G}}
		=
		\bigoplus_{a=1}^{n_M}
		\begin{bmatrix}
			V_{\rm ee}I_2 & \sqrt{V_{\rm ee}^2-1}Z\\
			\sqrt{V_{\rm ee}^2-1}Z & V_{\rm ee}I_2
		\end{bmatrix} .
	\end{aligned}
\end{equation}
Here, $V_{\rm ee}=1+v_{\rm el}/(1-\eta_e)$ is chosen to reproduce the trusted electronic noise in the detector model. $\mathrm{F}_0$ is introduced into the measured block through $n_M$ splitters with transmittance $\eta_e$ to obtain the detected block. Ultimately, the measurement collapses the unmeasured quantum state conditioned on an $n_M$-mode heterodyne outcome. The density matrix after conditioning is too complex to present here; please derive it according to Ref. \cite{laudenbach2018continuous}. Under quantum attacks, Eve's information is determined from the Holevo leakage of the corresponding covariance-matrix partition. Due to the leakage of classical information caused by the state broadcasting method, the impact of residual classical correlations is also considered through $I_{\mathrm{res}}$. In the following calculation, the classical mutual information is evaluated from the Gaussian covariance matrix of the heterodyne measurement outcomes derived from the quantum covariance matrix and the receiver model.

Under the four security levels $s\in\{\mathrm{AC},\mathrm{FS},\mathrm{CS},\mathrm{FC}\}$, namely the asymptotic case, finite-size effect, composable security, and finite-size composable security \cite{leverrier2010finite, leverrier2015composable}, the single-user SKR, the aggregate network SKR, and the all-measured joint network SKR of TS-QAN are formulated as
\begin{equation}
	\scalebox{0.99}
	{$
	\begin{aligned}
		K_{i^{\prime}}^{(s)}
		&=F_r f_s
		\left[\beta_{i^{\prime}} I_{\mathrm{AB},i^{\prime}}^{(s)}
		-\max\left(\chi_{\mathrm{EB},i^{\prime}}^{(s)}, I_{\mathrm{res},i^{\prime}}^{(s)}\right)
		-\Delta_{i^{\prime}}^{(s)}\right]_+,\\
		K_{\mathrm{net}}^{(s,\mathrm{sum})}
		&=\sum_{i^{\prime}=1}^{N}K_{i^{\prime}}^{(s)},\\
		K_{\mathrm{net}}^{(s,\mathrm{joint})}
		&=F_r f_s
		\left[\beta_{\mathrm{eff}}^{(s)}I_{\mathrm{AB},\mathrm{all}}^{(s)}
		-\chi_{\mathrm{EB},\mathrm{all}}^{(s)}
		-q_s\Delta^{(s)}\right]_+ .
	\end{aligned}
	$}
	\label{eq10}
\end{equation}
Here, $[x]_+=\max\left(x,0\right)$, $F_r$ is the repetition rate, $f_s$ is the post-processing prefactor of the selected security level, $N=N_WN_B$ is the network capacity, $I_{\mathrm{AB},i^{\prime}}^{(s)}$ is the useful Alice--Bob mutual information of User $i^{\prime}$, $\chi_{\mathrm{EB},i^{\prime}}^{(s)}$ is Eve's Holevo information, $I_{\mathrm{res},i^{\prime}}^{(s)}$ is the residual classical correlation between this user and the other Bob-side systems, and $\Delta_{i^{\prime}}^{(s)}$ is the finite-size or composable correction. The aggregate network SKR is obtained by summing independent user keys, whereas the all-measured policy uses the joint quantities $I_{\mathrm{AB},\mathrm{all}}^{(s)}$ and $\chi_{\mathrm{EB},\mathrm{all}}^{(s)}$. The correction multiplier is $q_s=1$ for a single-user partition and $q_s=N_W$ for the joint all-measured partition. The SKR per user is limited by the Pirandola-Laurenza-Ottaviani-Banchi (PLOB) bound \cite{pirandola2017fundamental}, while the network SKR is limited by the PLOB-$N$ \cite{pirandola2019bounds, pirandola2019end, das2021universal} bound.

\begin{figure*}[!h]
	\centering
	\subfigure[]{\label{fig5a}\includegraphics[width=1\linewidth]{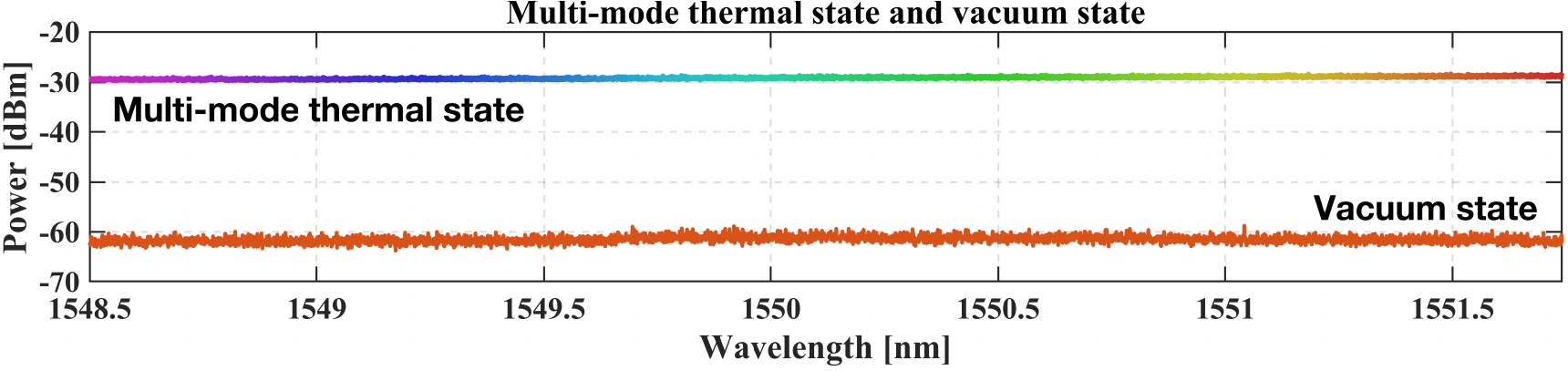}}
	\subfigure[]{\label{fig5b}\includegraphics[width=1\linewidth]{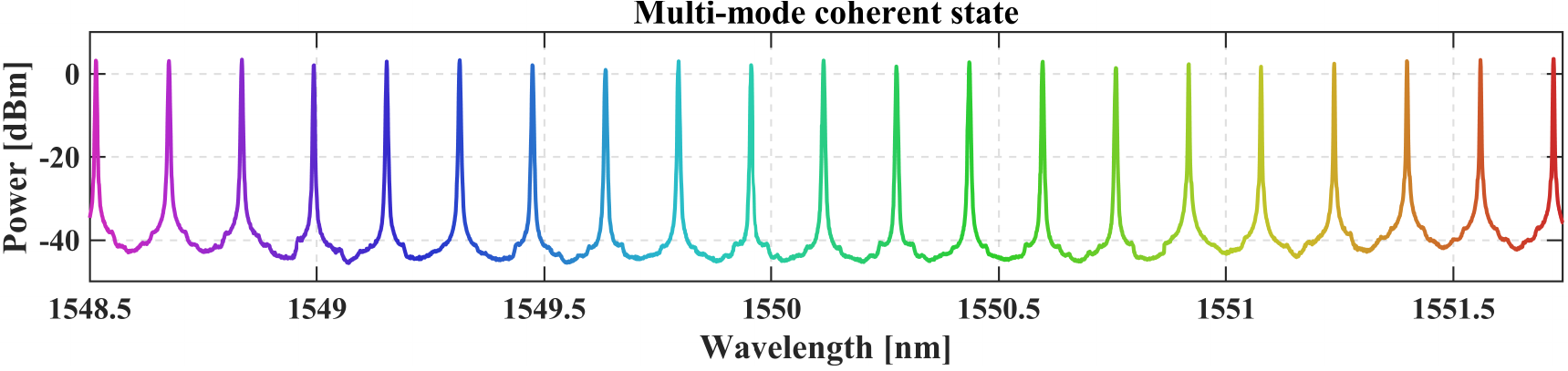}}
	\subfigure[]{\label{fig5c}\includegraphics[width=1\linewidth]{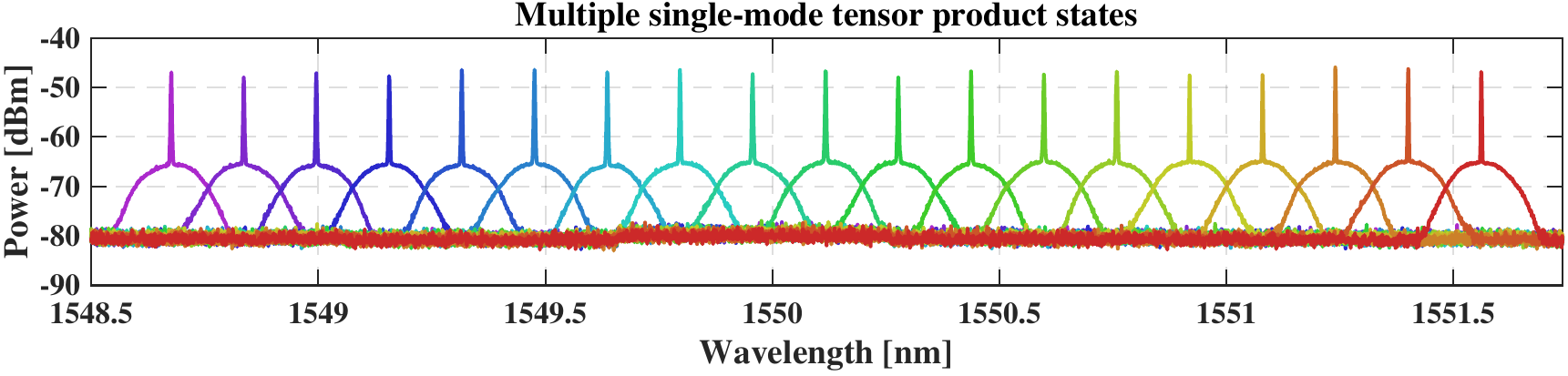}}
	\caption{Spectrum measurement of TS-QAN. (a) Multimode thermal state and vacuum state. When it was measured, the spectrometer operated in high-power setting with a minimum measurable power of $-60$ $\mathrm{dBm}$. (b) Multimode coherent state. The resolution of the spectrometer is $1.8$ $\mathrm{pm}$. (c) Multiple single-mode tensor product states. When it was measured, the spectrometer operated in low-power setting with a minimum measurable power of $-80$ $\mathrm{dBm}$.}
	\label{fig5}
\end{figure*}

\begin{figure*}[!h]
	\centering
	\subfigure[]{\label{fig6a}\includegraphics[width=1\linewidth]{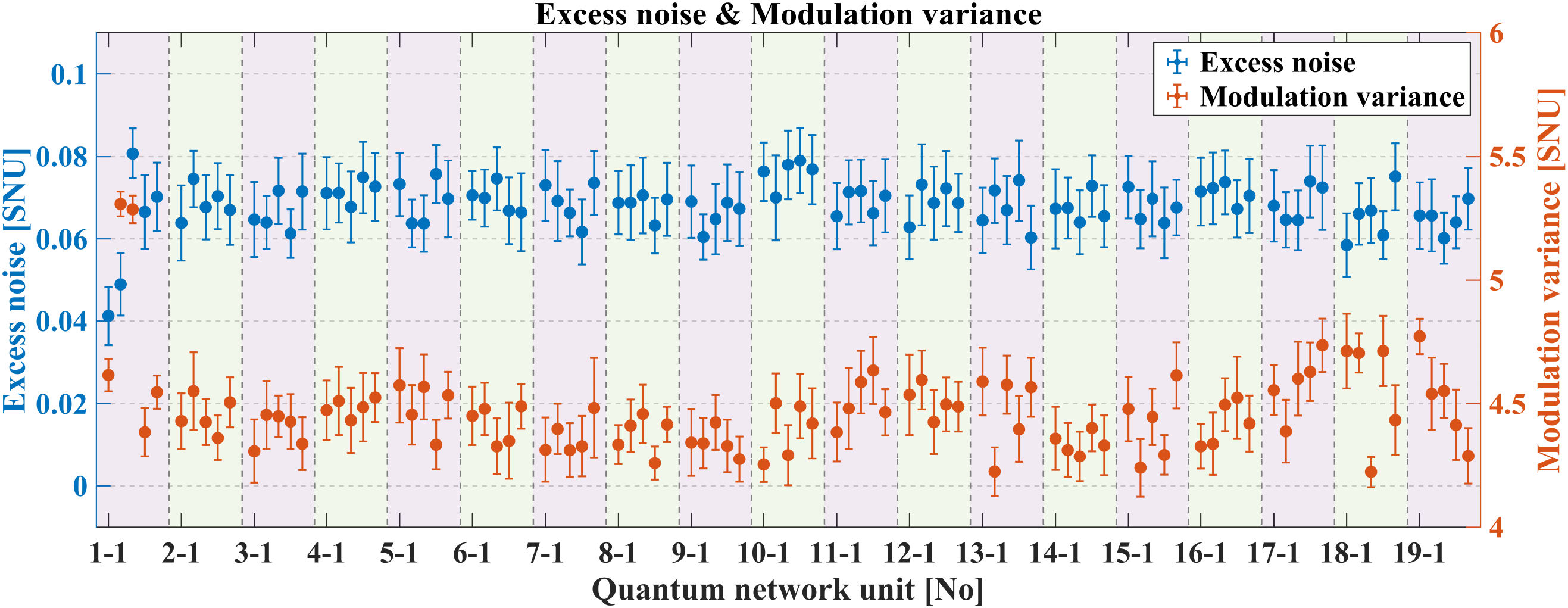}}
	\subfigure[]{\label{fig6b}\includegraphics[width=1\linewidth]{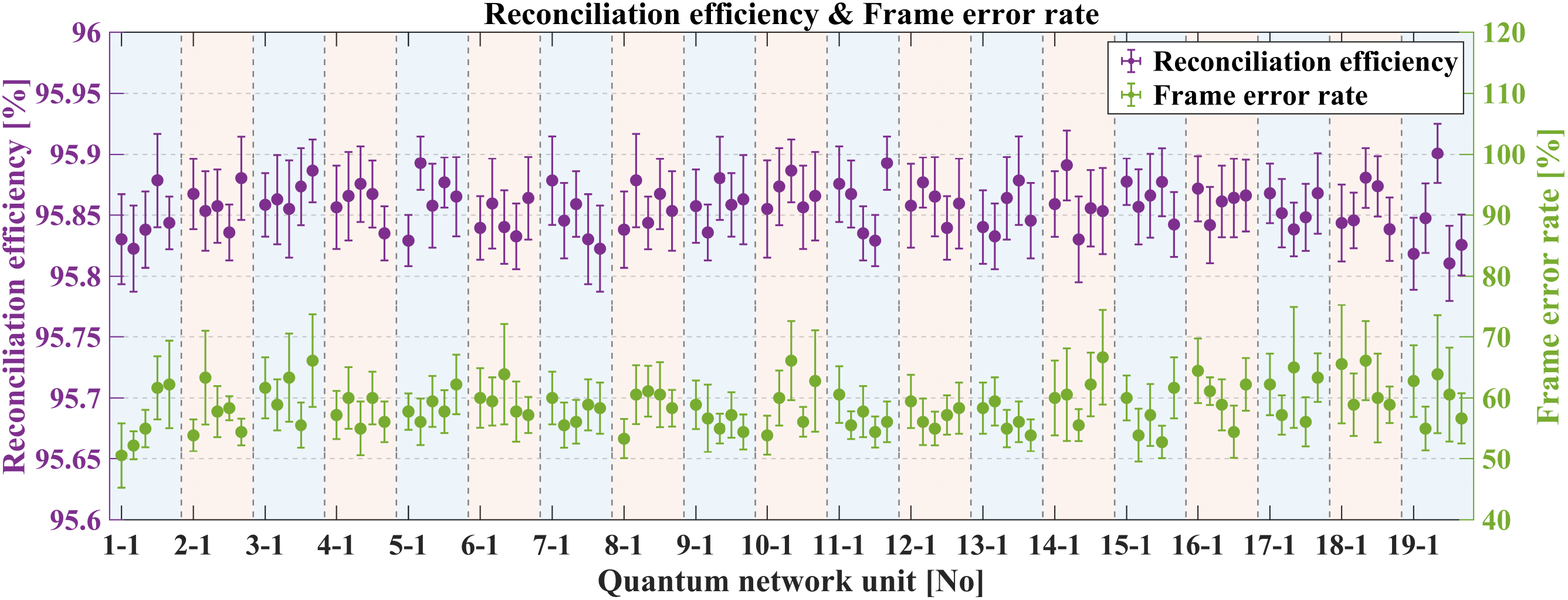}}
	\caption{Parameter estimation of TS-QAN. (a) Excess noise and modulation variance. The blue scatter points correspond to the excess noise on the left. The red scatter points correspond to the modulation variance on the right. (b) Reconciliation efficiency and frame error rate. The purple scatter points correspond to the reconciliation efficiency on the left. The green scatter points correspond to the frame error rate on the right.}
	\label{fig6}
\end{figure*}

\begin{figure*}[!h]
	\centering
	\subfigure[]{\label{fig7a}\includegraphics[height=8.15cm]{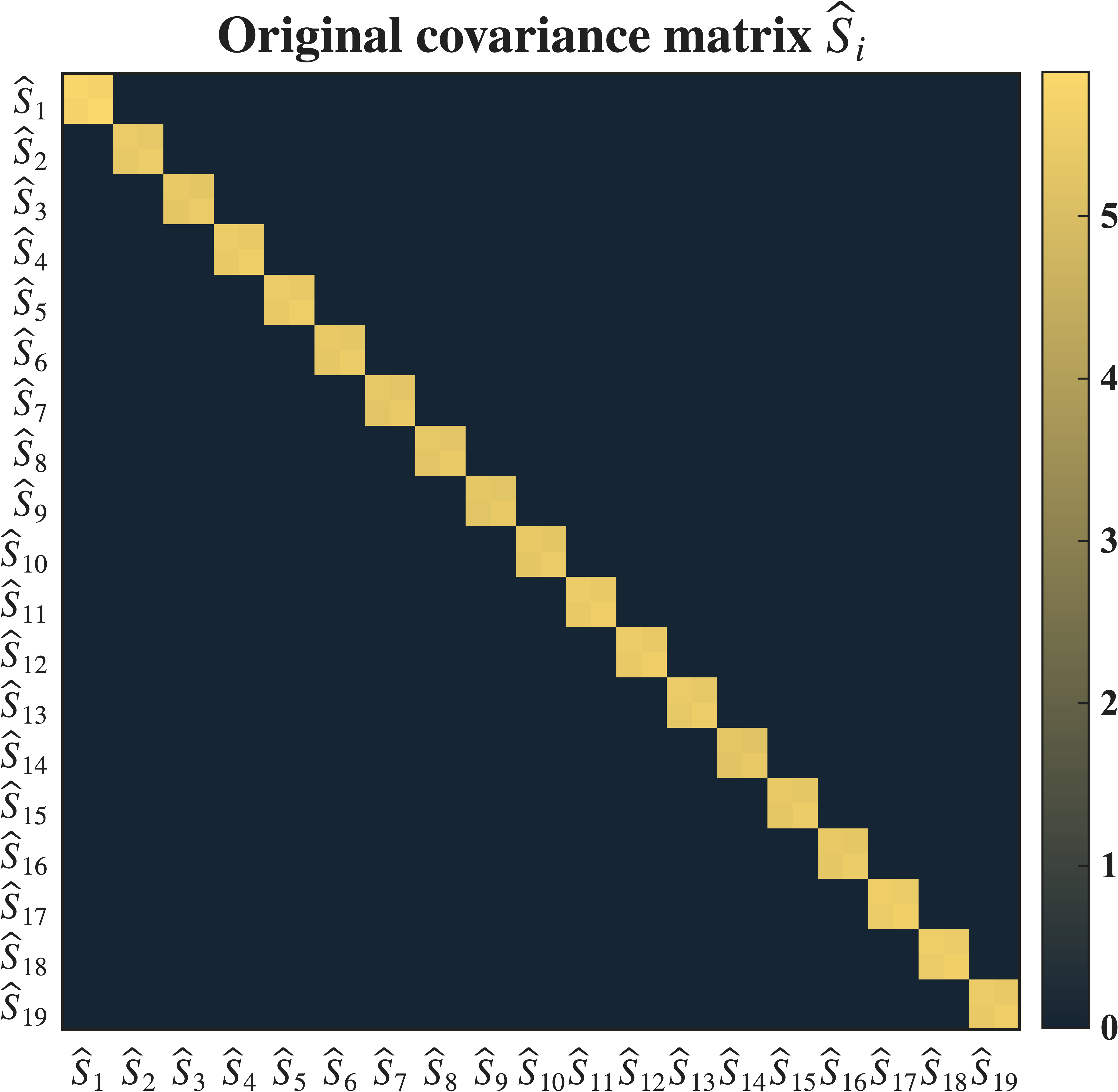}}
	\subfigure[]{\label{fig7b}\includegraphics[height=8.15cm]{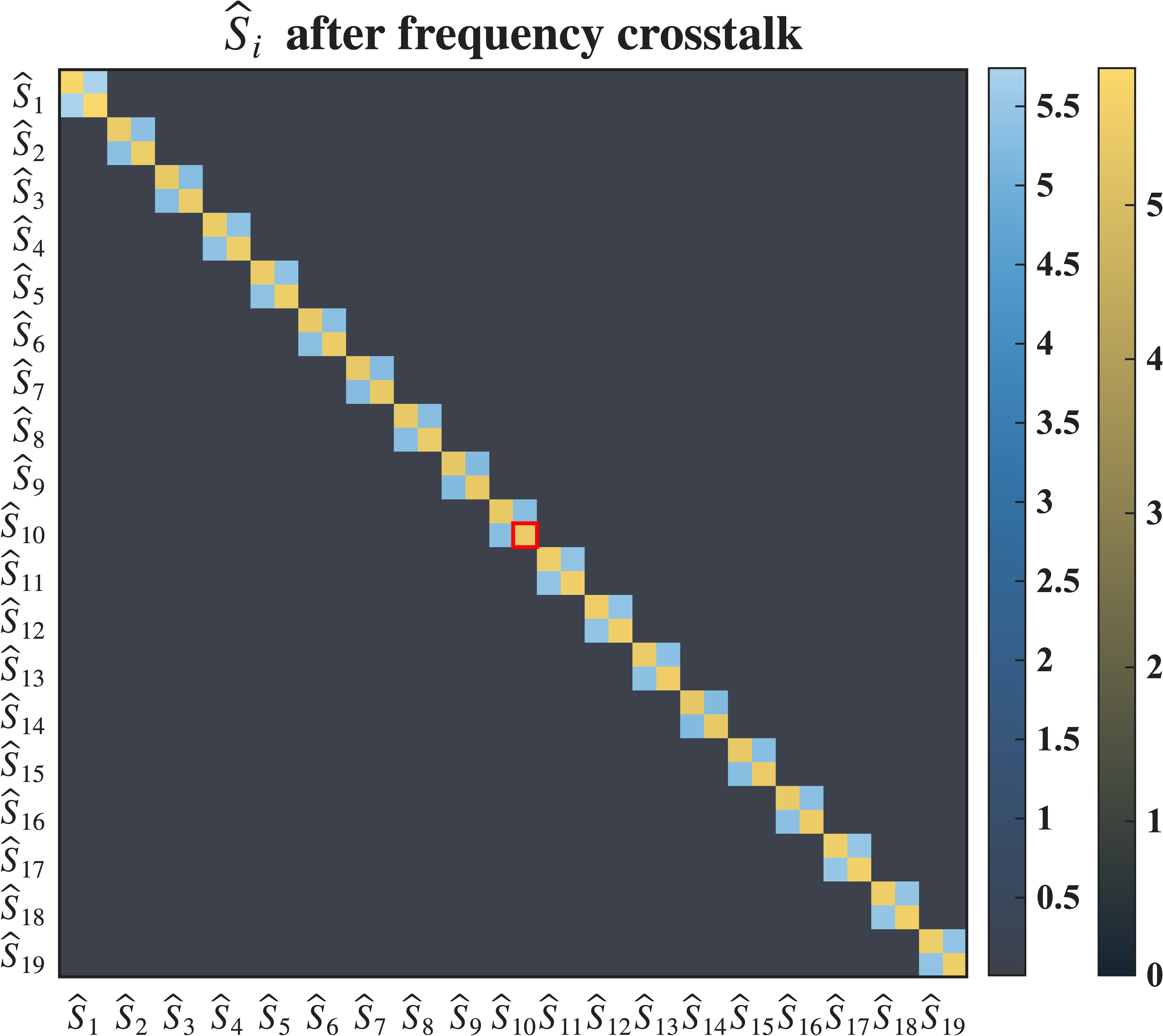}}
	\subfigure[]{\label{fig7c}\includegraphics[height=7.73cm]{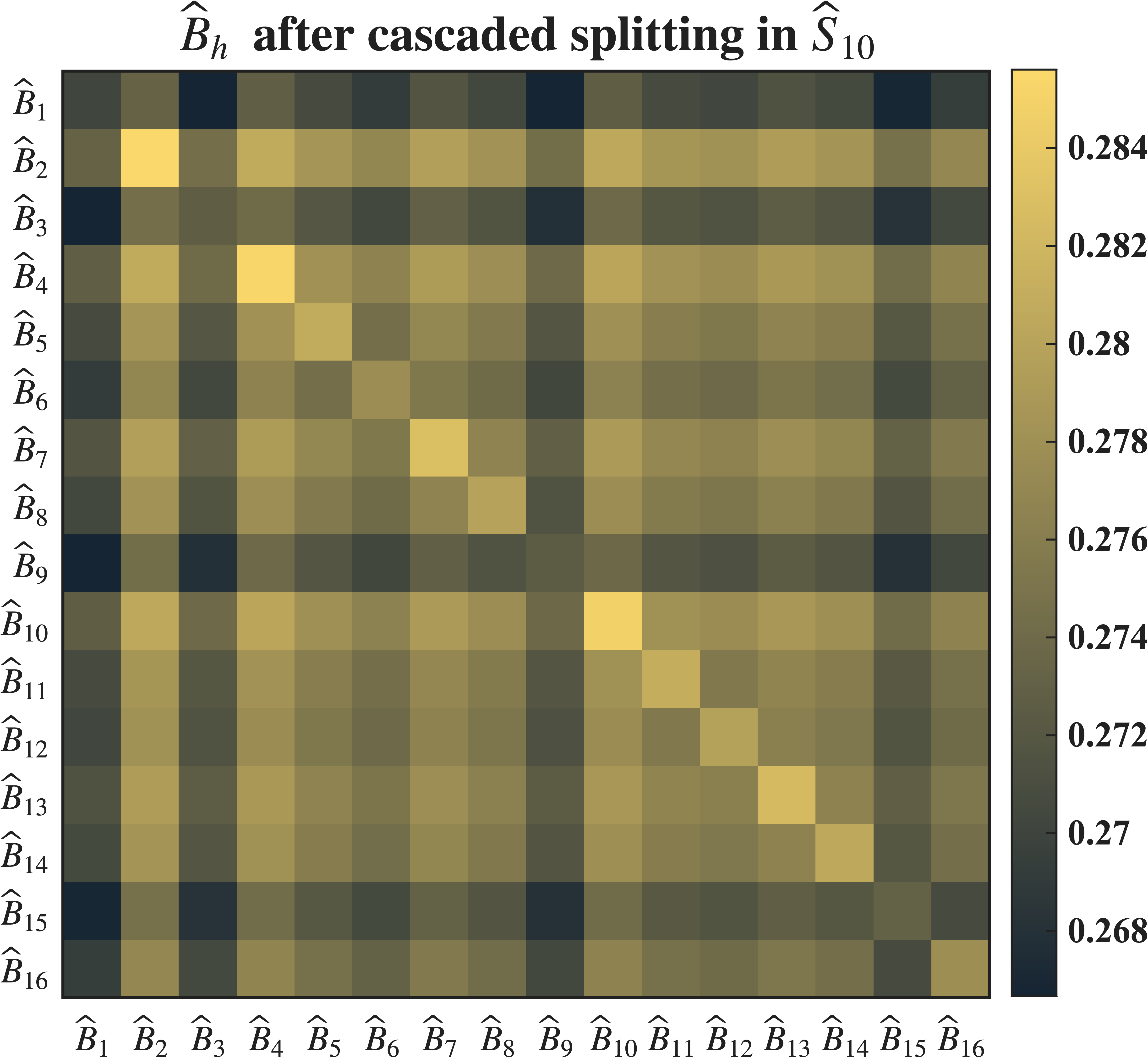}}
	\subfigure[]{\label{fig7d}\includegraphics[height=7.73cm]{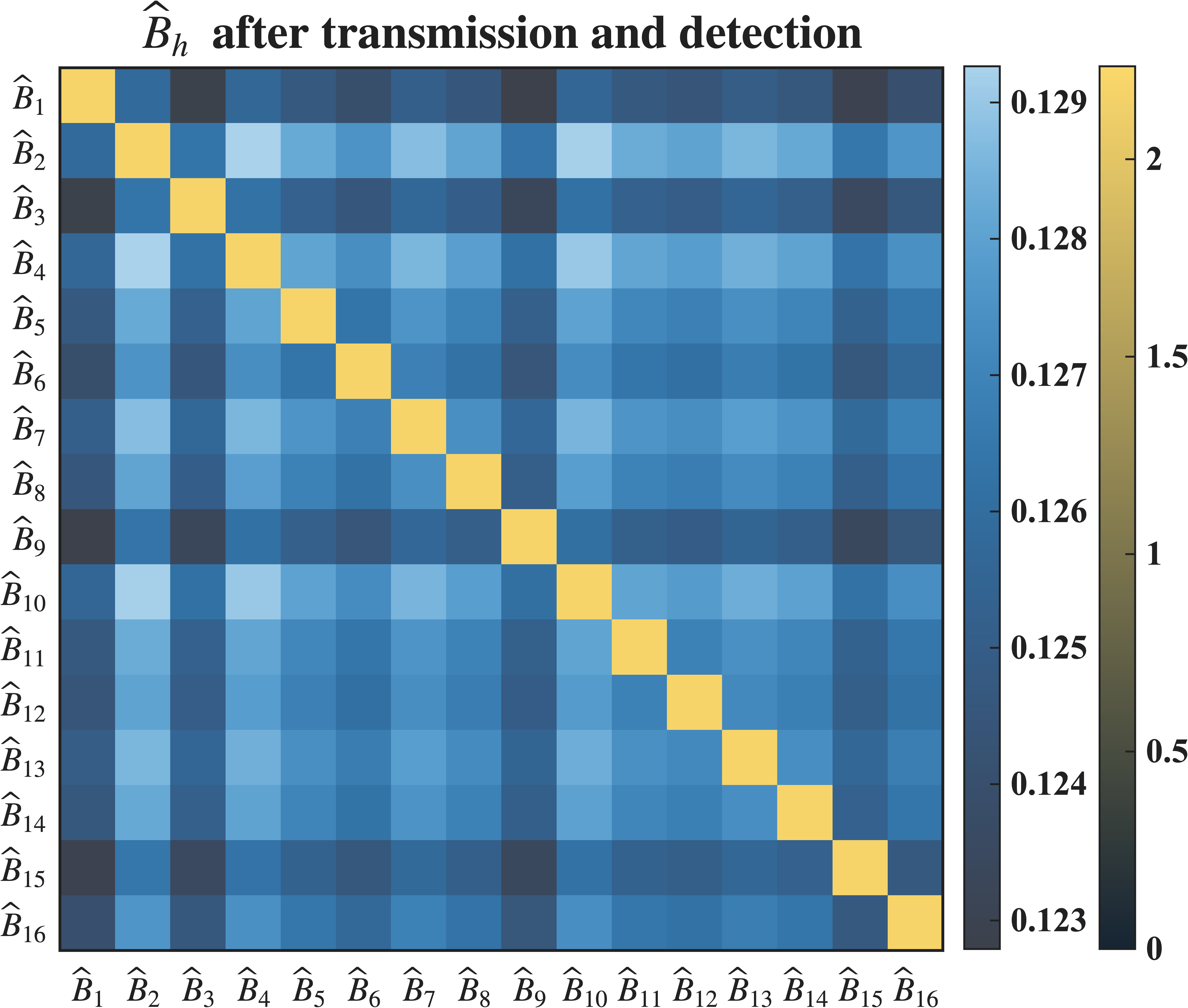}}
	\caption{Matrix depiction of TS-QAN. The matrices are reconstructed from pairwise measurements under stable and calibrated operating conditions, rather than acquired through a single simultaneous full-network covariance measurement. (a) Initial Alice--Bob covariance block before frequency allocation, showing the diagonal EPR-pair structure. (b) Alice--Bob covariance block after frequency allocation, showing the non-diagonal correlations induced by finite mode isolation. (c) Representative Alice–Bob covariance slice after frequency allocation and cascaded splitting, illustrating the user-level correlation pattern within the reconstructed $19\times304$ structure. (d) Representative Bob-side branch covariance block, illustrating the branch-network correlations embedded in the full $304\times304$ Bob-side covariance matrix.}
	\label{fig7}
\end{figure*}

\begin{figure*}[!h]
	\centering
	\subfigure[]{\label{fig8a}\includegraphics[width=0.49\linewidth]{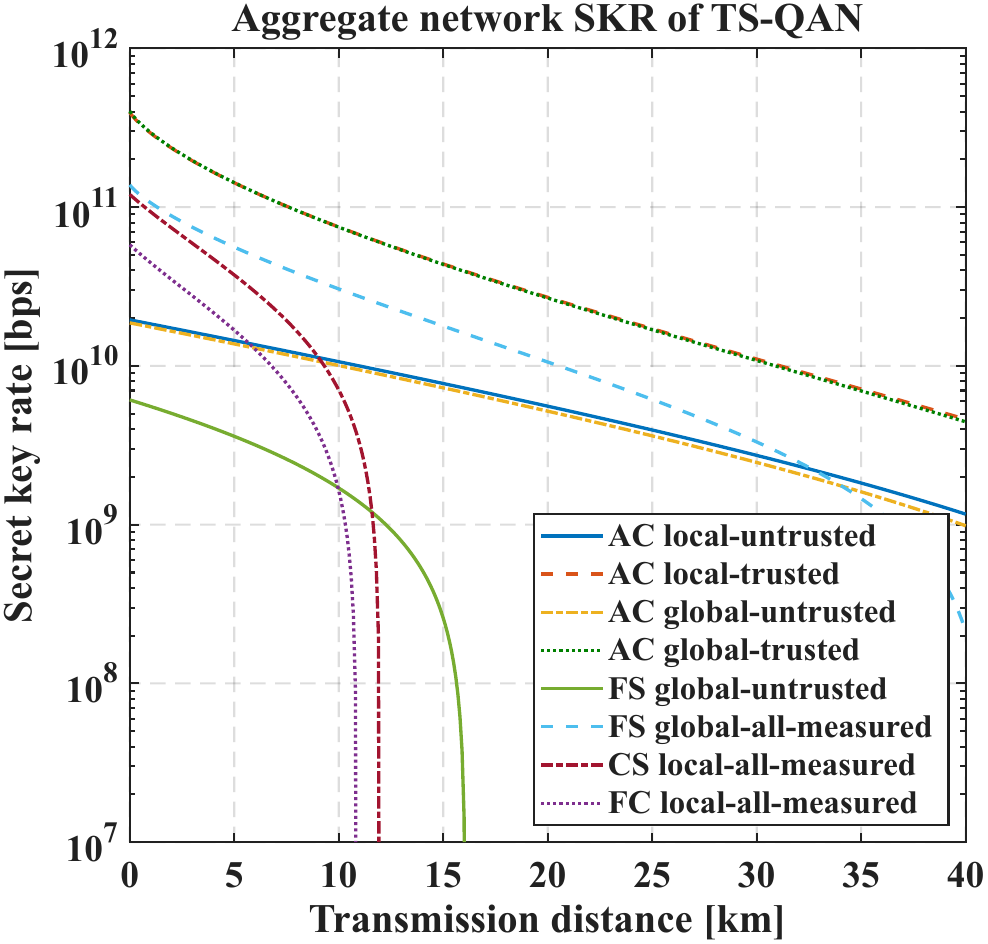}}
	\subfigure[]{\label{fig8b}\includegraphics[width=0.49\linewidth]{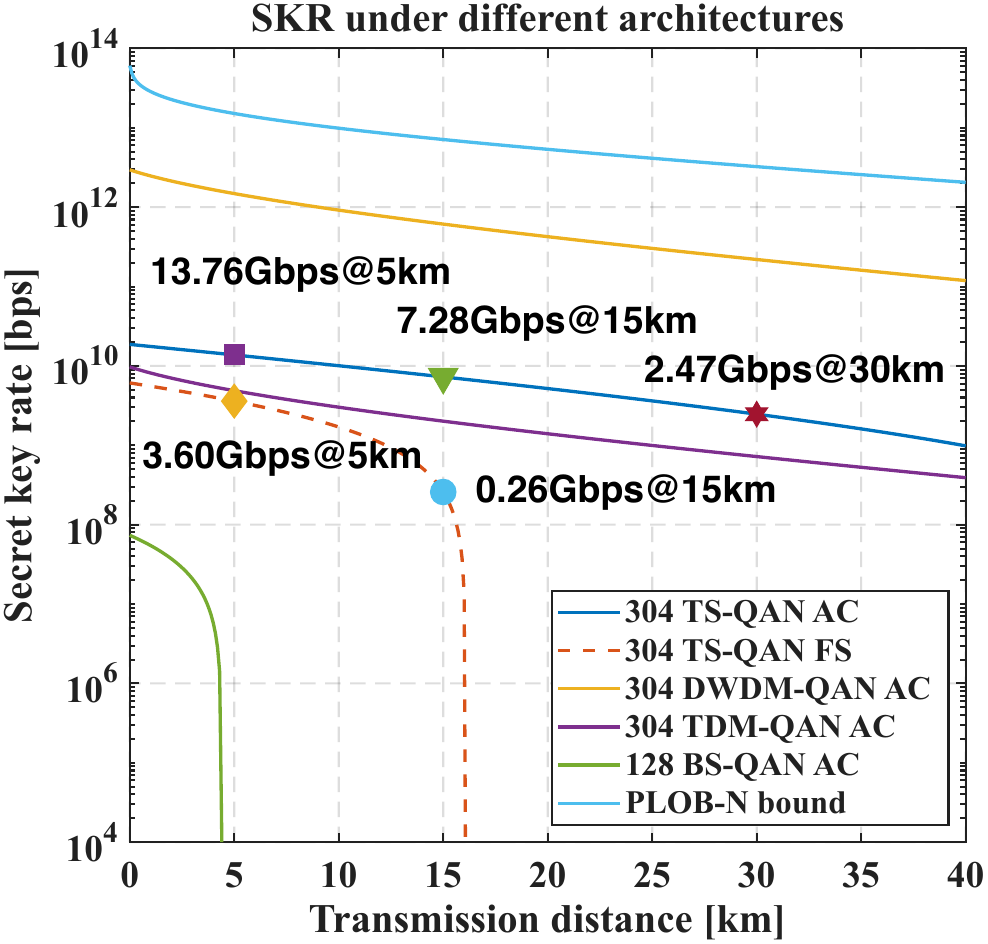}}
	\subfigure[]{\label{fig8c}\includegraphics[width=1\linewidth]{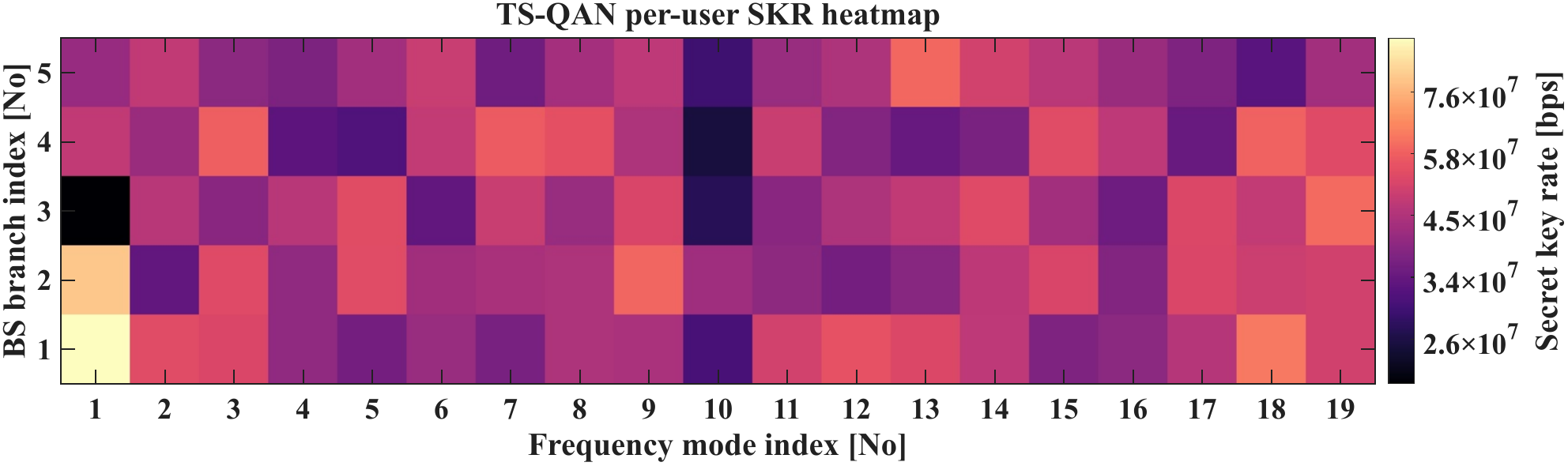}}
	\caption{Performance analysis of TS-QAN. (a) Experimental aggregate network SKR under different receiver-security models. These representative curves are selected from the full organization of two receiver capabilities, three security partitions, and four security levels: the receiver capability is either local or global; the security partition is untrusted, trusted, or all-measured; and the security level is asymptotic case, finite-size effect, composable security, or finite-size composable security. (b) Aggregate network SKR comparison among different architectures. The five experimental TS-QAN results are included in the comparison: AC gives $13.76$ $\mathrm{Gbps}$ over $5$ $\mathrm{km}$, $7.28$ $\mathrm{Gbps}$ over $15$ $\mathrm{km}$, and $2.47$ $\mathrm{Gbps}$ over $30$ $\mathrm{km}$, while FS gives $3.60$ $\mathrm{Gbps}$ over $5$ $\mathrm{km}$ and $0.26$ $\mathrm{Gbps}$ over $15$ $\mathrm{km}$. (c) Experimental SKR heatmap per sampled user. $5$ users are sampled from each of the $19$ frequency modes, and the heatmap shows the reconstructed branch-dependent SKR distribution of these sampled users.}
	\label{fig8}
\end{figure*}

\subsection*{Experimental Setup}

In this section, we will introduce the experimental setup of TS-QAN. As depicted in Fig. \ref{fig4}, the experimental setup can be divided into three parts: QLT, channel, and QNUs.

\emph{\textbf{QLT}} ----- The QLT consists of $1$ Amplified Spontaneous Emission (ASE) source, $1$ comb laser, $2$ WaveShapers (WSs), and $19$ Passive Preparation Modules (PPMs). The ASE source has a spectrum center at $1550$ $\mathrm{nm}$ and a bandwidth exceeding $5$ $\mathrm{THz}$. The comb laser is a home-built EO comb centered at $1550$ $\mathrm{nm}$ with a $20$ $\mathrm{GHz}$ spacing. The multimode thermal state emitted by the ASE source and the multimode coherent state emitted by the comb laser are fed into $2$ $1 \times 19$ WSs with a mode isolation of $30$ $\mathrm{dB}$, respectively. The function of WS is to separate a multimode state into multiple single-mode states (without considering mode crosstalk), with each filtering window of $20$ $\mathrm{GHz}$. By adjusting the position of the filtering windows, the centers of each single-mode thermal state and each single-mode coherent state are aligned. Then, $19$ pairs of single-mode thermal state and coherent state of the same color are input into $19$ PPMs. Each PPM comprises a Fiber Polarizer (FP), a $10:90$ $1 \times 2$ BS, a $1:99$ $1 \times 2$ BS, a $50:50$ $2 \times 1$ BS, a Variable Optical Attenuator (VOA), and an Integrated Coherent Receiver (ICR). $\mathrm{INPUT-}1$ port and $\mathrm{INPUT-}2$ port receive the single-mode thermal state and the single-mode coherent state, respectively. The thermal state is transformed from circular polarization to linear polarization through the FP. Next, it is split by the $1:99$ $1 \times 2$ BS into a high-power part and a low-power part. The high-power thermal state enters the ICR as the original signal. The low-power thermal state is directed into the $50:50$ $2 \times 1$ BS. The coherent state is split by the $10:90$ $1 \times 2$ BS into a high-power part and a low-power part. The high-power coherent state enters the ICR as the QLT's LO. The low-power coherent state is directed into the $50:50$ $2 \times 1$ BS. The ICR performs heterodyne detection to obtain the QLT's key, with an electronic noise of $0.31$ $\mathrm{SNU}$ and a quantum detection efficiency of $56\%$. The output from the $50:50$ $2 \times 1$ BS is composed of the single-mode thermal state centered with the single-mode coherent state serving as the beacon, which is a monochromatic tensor product state. Before leaving the $\mathrm{OUTPUT}$ port, it enters the VOA for appropriate power adjustment. Upon exiting the PPMs, polychromatic tensor product states are sent into the channel.

\emph{\textbf{Channel}} ----- The channel consists of $19$ Cascaded BS (CBS) and $304$ single-mode fiber cables. Each $1 \times 16$ CBS receives a monochromatic tensor product state. The $19$ tensor product states are each split into $16$ equal parts, resulting in $304$ tensor product states. Subsequently, they are transmitted to $304$ QNUs via $304$ single-mode fiber cables with an attenuation coefficient of $0.17$ $\mathrm{dB/km}$.

\emph{\textbf{QNUs}} ----- Each QNU consists of $1$ single-frequency laser, $1$ Polarization Controller (PC), and $1$ ICR. The QNU's LO is generated by the single-frequency laser. The polarization of the single-mode tensor product state from the $\mathrm{INPUT}$ port is adjusted by the PC to align with that of the LO. Afterwards, the tensor product state and the LO are fed into the ICR for heterodyne detection. The QNU utilizes DSP algorithms to achieve optimal reception of quantum signals. The steps of the DSP algorithm are as follows: Filter the beat signal converted from the beacon tone; Calculate the time-varying phase according to the beat signal; Perform time-varying phase recovery on the quantum signal; Suppress the residual near-direct-current beat component by direct-current-offset removal; Downsample the quantum signal to obtain the raw data. Following the execution of DSP, parameters such as excess noise, modulation variance, and covariance matrix are measured and estimated. The same DSP pipeline is used consistently in signal acquisition, shot-noise calibration, and electronic-noise calibration, so that the measured quantities correspond to the same effective detection mode, as detailed in the Supplementary Information. Lastly, data post-processing is conducted. Through reverse reconciliation, each QNU and the QLT generate independent keys. The Multi-Edge-Type Low-Density Parity-Check (MET-LDPC) codes \cite{zhang2025effective}, improved by whitening noise \cite{zhang2020long}, are employed for decoding to get the reconciliation efficiency and frame error rate. Privacy amplification is carried out through hash compression, ensuring that the amount of key held by each user remains within the secure bound.

\subsection*{Experimental Results}

In this section, we will present the experimental results of TS-QAN. As illustrated in Figs. \ref{fig5}, \ref{fig6}, \ref{fig7}, and \ref{fig8}, the experimental results can be divided into four parts: spectrum measurement, parameter estimation, matrix depiction, and performance analysis.

\emph{\textbf{Spectrum Measurement}} ----- The results of the spectrum measurement are shown in Fig. \ref{fig5}. Here, we present the measurement results of spectra for the three states: thermal state, coherent state, and tensor product state. The spectra of the multimode thermal state and the vacuum state are displayed as the gradient-colored and red lines in Fig. \ref{fig5a}, respectively. Since the bandwidth of the ASE source is much larger than the bandwidth selected in the experiment, the spectrum of the thermal state is very flat. Its flatness is $1.24$ $\mathrm{dB}$, and the average power is $-29.10$ $\mathrm{dBm}$. The spectrum of the multimode coherent state is illustrated as the gradient-colored line in Fig. \ref{fig5b}. Due to the limited resolution of the spectrometer, the linewidth of the coherent state is difficult to observe directly. Its linewidth is less than $1$ $\mathrm{kHz}$, and the flatness is $2.63$ $\mathrm{dB}$. The spectra of multiple single-mode tensor product states are depicted as several solid-colored lines with different colors in Fig. \ref{fig5c}. Because practical filters cannot achieve an ideal frequency response, the visible bandwidth of the tensor product states is greater than the filtering window. Their $3$ $\mathrm{dB}$ bandwidth is approximately $20$ $\mathrm{GHz}$, and the beacon exceeds the signal by about $20$ $\mathrm{dBm}$.

\emph{\textbf{Parameter Estimation}} ----- The results of the parameter estimation are shown in Fig. \ref{fig6}. Here, we present the estimation results of $95$ sampled users for the four parameters: excess noise, modulation variance, reconciliation efficiency, and frame error rate. The sampling is performed by randomly selecting $5$ users from each color for testing. The block size is ${10}^{10}$ per user. The estimation results for excess noise and modulation variance are displayed as blue and red scatter points in Fig. \ref{fig6a}, respectively. Due to the passive preparation structures and high repetition rates, the excess noise in thermal state networks is slightly higher than that in coherent state networks. The main practical source of the additional excess noise in the present implementation is the high-bandwidth operation of the system, including dispersion-related impairment and bandwidth-limited receiver noise, rather than the passive-preparation mechanism itself. Benefiting from the flat spectrum of the thermal state, the equal frequency slicing, the equal power splitting, and the power adjustment before transmission, the modulation variance can be approximately equalized across wavelength channels, although practical imperfections still result in residual channel-to-channel deviations. The mean value of excess noise is $0.07$ $\mathrm{SNU}$, and the mean value of modulation variance is $4.47$ $\mathrm{SNU}$. The estimation results for reconciliation efficiency and frame error rate are plotted as purple and green scatter points in Fig. \ref{fig6b}, respectively. Thanks to the MET-LDPC code improved by whitening noise, the decoding performance of the thermal state network approaches the theoretical limit. The mean value of reconciliation efficiency is $95.86\%$, and the mean value of frame error rate is $58.70\%$.

\emph{\textbf{Matrix Depiction}} ----- The results of the matrix depiction are shown in Fig. \ref{fig7}. Here, we display the quantum covariance matrices associated with the EB model. In this matrix depiction, $\hat{S}_{i}$ is used as a compact schematic label for the Alice--Bob quadrature-operator pair associated with a displayed covariance subblock, namely $\hat{S}_{i}\equiv(\hat{A}_{i},\hat{B}_{i'})$, where $\hat{A}_{i}=(\hat{X}^{i}_{A},\hat{P}^{i}_{A})$ denotes the Alice-side quadrature operators of the $i$-th frequency mode and $\hat{B}_{i'}=(\hat{X}^{i'}_{B},\hat{P}^{i'}_{B})$ denotes the Bob-side quadrature operators of the $i'$-th user mode. Fig. \ref{fig7a} depicts the initial Alice--Bob covariance block before frequency allocation, where only the diagonal EPR-pair correlations are present. Fig. \ref{fig7b} depicts the Alice--Bob covariance block after the frequency-allocation stage. Finite mode isolation makes the off-diagonal elements of $D_{\rm full}$ nonzero beyond the ideal diagonal terms, so the Alice--Bob covariance block becomes non-diagonal. Fig. \ref{fig7c} depicts the user-level Alice--Bob covariance block after frequency allocation and cascaded splitting. The $N_W$ frequency modes are expanded to $N_WN_B$ Bob user modes, so the matrix directly reflects the $19\times304$ Alice--user correlation structure. Fig. \ref{fig7d} depicts the Bob-side user covariance block. Different BS branches generally retain nonzero quantum covariance because they share a common input mode, while the independent vacuum modes introduced by the transmission and detection contribute to the diagonal vacuum-noise terms. Therefore, the Bob-side quantum covariance matrix is generally non-diagonal rather than block diagonal.

\emph{\textbf{Performance Analysis}} ----- The results of the performance analysis are shown in Fig. \ref{fig8}. Here, the performance is presented from three aspects: aggregate network SKR under representative receiver-security models, aggregate SKR comparison among different network architectures, and the reconstructed per-user SKR distribution of sampled users. Fig. \ref{fig8a} shows the experimental network SKR of TS-QAN under representative combinations of receiver capability, security partition, and security level. The curves compare different receiver capabilities, security partitions, and security levels using the same reconstructed TS-QAN covariance matrix. The local receiver keeps only the main frequency component available to each QNU, whereas the global receiver uses the frequency-resolved components arriving at the same QNU. The untrusted, trusted, and all-measured partitions correspond to different treatments of receiver-side components and measured correlations in the security evaluation. In the present TS-QAN, each monochromatic state is further distributed through a cascaded beam-splitting structure, which significantly reduces the effective transmittance of each user branch. Consequently, the additional composable and finite-size-composable corrections become severe under the current experimental block size. Nevertheless, when receiver-side information is incorporated through the all-measured security policy, residual correlations can be more explicitly accounted for in the covariance-matrix analysis, allowing positive composable and finite-size composable SKRs to be obtained in the present parameter regime, although they remain lower than the corresponding AC and FS results. In addition to the receiver-security comparison, Fig. \ref{fig8b} compares the aggregate network SKR of TS-QAN with different access-network architectures under the same aggregate-network metric, including DWDM-QAN with $304$ users, Time-Division-Multiplexing QAN (TDM-QAN) with $304$ users, BS-QAN with $128$ users, and the PLOB-$N$ bound. The five experimental TS-QAN aggregate SKR points are included in Fig. \ref{fig8b}: AC gives $13.76$ Gbps over $5$ km, $7.28$ Gbps over $15$ km, and $2.47$ Gbps over $30$ km, while FS gives $3.60$ Gbps over $5$ km and $0.26$ Gbps over $15$ km. The TS-QAN asymptotic result lies between the corresponding DWDM-QAN and TDM-QAN asymptotic references, while the finite-size TS-QAN result remains much higher than the asymptotic BS-QAN reference. Fig. \ref{fig8c} further shows the per-user SKR heatmap reconstructed from $95$ sampled users, with $5$ sampled users in each of the $19$ frequency modes. For the full $304$-QNU evaluation, the aggregate network SKR is calculated using the reconstructed $304$-QNU covariance matrix. Among the full $304$ QNUs, $95$ QNUs are long-time calibrated for trusted CV-QKD scalar parameters, while the remaining $209$ QNUs use the operating-point covariance entries together with conservative parameter filling from the corresponding calibrated frequency group, as detailed in the Supplementary Information. Because of finite mode isolation and state broadcasting, network-level and point-to-point evaluations can differ; the network evaluation retains the non-diagonal covariance structure and provides the network-aware conservative result adopted for the aggregate SKR.

\section*{Discussion}

In the presence of phase noise, the beacon beat component is first used for time-varying frequency-offset and phase-noise estimation and compensation \cite{xu2024robust, xu2025ofdm, xu2026polychromatic}. After compensation, the residual beat contribution is largely converted into a narrowband near-direct-current term and is further suppressed by direct-current-offset removal. Due to residual phase noise this suppression is not mathematically perfect; however, the remaining contribution is subdominant in our operating regime and is accounted for in the experimentally estimated noise parameters. Importantly, we explicitly treat these DSP steps as an integral part of the optimal receiver model. As emphasized in Ref. \cite{hajomer2022modulation}, digital filtering operations define the effective measured temporal mode and must therefore be specified and applied consistently, while the filter design should avoid introducing intersymbol interference or additional information leakage. Following this viewpoint, we use the same DSP pipeline throughout the calibration and data-processing stages. This is in line with recent continuous-mode treatments of DSP in CV-QKD security analysis \cite{chen2023continuous}.

In the present article, the channel excess noise is modeled as Alice-referred excess noise for each user channel, while network effects are incorporated through finite mode isolation and the resulting non-diagonal covariance structure. Therefore, the current analysis should be understood as a physically motivated network Gaussian-attack model, rather than as an exhaustive optimality proof over all possible independent or correlated excess-noise injections in passive optical networks. A more general passive-network attack model with explicitly correlated excess-noise covariance across modes can in principle be formulated within the same covariance-matrix framework, but this extension is beyond the scope of the present proof-of-principle research.

We note that the finite-size and composable terms appearing in Eq. \eqref{eq10} are adopted from point-to-point CV-QKD analyses and are used here as indicative extensions for discussing performance penalties in the network setting. The main rigorous security result established in the present article is the asymptotic covariance-matrix-based network analysis under mode crosstalk and state-broadcasting-induced correlations, rather than a full finite-size composable security proof for the network protocol. A complete finite-size composable security proof for the present network protocol, fully adapted to the multimode and broadcast network structure, is beyond the scope of this article and remains an important direction for future work. Accordingly, although an optimized active-preparation system may yield a larger finite-size key under some operating regimes, the present results should not be interpreted as evidence of a fundamental sub-optimality of passive preparation. The main objective of TS-QAN is architecture-level scalability through shared-resource distribution, rather than single-link finite-size dominance.

Even as the network capacity scales to hundreds of users, TS-QAN still delivers impressive SKR. Assuming all QANs can achieve the same repetition rate, TS-QAN surpasses all other QANs except DWDM-QAN. In practice, active-preparation QANs require parallel high-speed modulation, driving, calibration, and DSP chains as the number of user channels increases, so their network-level exploitation of high-bandwidth quantum randomness is limited by system-scaling complexity rather than by modulator bandwidth alone. BS-QAN has the simplest architecture, but its performance is far from meeting the needs of modern communication. TDM-QAN offers acceptable performance, but there are no optical switches available to support hundreds of users. DWDM-QAN exhibits exceptional performance, but hundreds of polychromatic resources are prohibitively complex and expensive. In contrast, the core components of TS-QAN, such as the thermal source and the flat optical comb, are cost-effective and widely available devices with numerous mature solutions for miniaturization and integration. This makes TS-QAN a promising solution under the current trade-off among performance, scalability, and implementation complexity.

In the experiment, we utilized only a negligible $380$ $\mathrm{GHz}$ bandwidth out of the total $5$ $\mathrm{THz}$ bandwidth. Using merely $7.6\%$ of the total bandwidth, TS-QAN has already met the requirements of advanced and dominant CAN. In the future, we will further optimize the structure and leverage more bandwidth to build a quantum communication network with larger capacity and higher rate. This will not only provide high-performance network solutions for quantum communications but also offer seamless interconnection between quantum computers in the upcoming era of the quantum internet.

\section*{Conclusion}

To satisfy the OTP encryption demands of advanced and dominant CAN, we proposed TS-QAN. Firstly, a theoretical analysis was conducted through the PM model and the EB model. Subsequently, an experimental platform was constructed and proof-of-principle experiments were performed. Finally, the experimental results aligned with the theoretical expectations, confirming that TS-QAN satisfies the specifications of advanced and dominant CANs. This work offers a promising QKD network architecture for commercial telecommunication networks under the current trade-off among performance, scalability, and implementation complexity.

\section*{Acknowledgments}

\subsection*{Author Contributions}

G. Z. conceived the research. Y. X., Q. Z., and G. Z. conducted the mathematical derivation. Y. X., Q. Z., X. L., Z. G., P. T., H. Y., and T. W. carried out the experiment. Y. X., X. L., and T. W. analyzed the data and wrote the manuscript. P. H. provided the technical guide for performance analysis and data post-processing. All authors contributed to the data collection, discussed the results, and reviewed the manuscript.

\subsection*{Funding}

This work was supported by the National Natural Science Foundation of China (Grant No. 62571316, 61971276), Quantum Science and Technology-National Science and Technology Major Project (Grant No. 2021ZD0300703), Shanghai Municipal Science and Technology Major Project (2019SHZDZX01), Natural Science Foundation of Shanghai (Grant No. 25ZR1402251), and Project supported by Cultivation Project of Shanghai Research Center for Quantum Sciences (Grant No. LZPY2024).

\subsection*{Competing Interests}

The authors declare that there is no conflict of interest regarding the publication of this article.

\subsection*{Data Availability}

The data that support the plots within this paper are available from the corresponding authors upon reasonable request.

\subsection*{Code Availability}

Computer codes to calculate the information capacity are available from the corresponding authors on reasonable request.

\bibliography{TSQAN.bib}	

\begin{thebibliography}{10}
\expandafter\ifx\csname url\endcsname\relax
  \def\url#1{\texttt{#1}}\fi
\expandafter\ifx\csname urlprefix\endcsname\relax\def\urlprefix{URL }\fi
\expandafter\ifx\csname doiprefix\endcsname\relax\def\doiprefix{DOI }\fi
\providecommand{\bibinfo}[2]{#2}
\providecommand{\eprint}[2][]{\url{#2}}

\bibitem{arute2019quantum}
\bibinfo{author}{Arute, F.} \emph{et~al.}
\newblock \bibinfo{journal}{\bibinfo{title}{Quantum supremacy using a
  programmable superconducting processor}}.
\newblock {\emph{\JournalTitle{Nature}}} \textbf{\bibinfo{volume}{574}},
  \bibinfo{pages}{505--510} (\bibinfo{year}{2019}).

\bibitem{rivest1978method}
\bibinfo{author}{Rivest, R.~L.}, \bibinfo{author}{Shamir, A.} \&
  \bibinfo{author}{Adleman, L.}
\newblock \bibinfo{journal}{\bibinfo{title}{A method for obtaining digital
  signatures and public-key cryptosystems}}.
\newblock {\emph{\JournalTitle{Communications of the ACM}}}
  \textbf{\bibinfo{volume}{21}}, \bibinfo{pages}{120--126}
  (\bibinfo{year}{1978}).

\bibitem{scarani2009security}
\bibinfo{author}{Scarani, V.} \emph{et~al.}
\newblock \bibinfo{journal}{\bibinfo{title}{The security of practical quantum
  key distribution}}.
\newblock {\emph{\JournalTitle{Reviews of modern physics}}}
  \textbf{\bibinfo{volume}{81}}, \bibinfo{pages}{1301--1350}
  (\bibinfo{year}{2009}).

\bibitem{wang2022sub}
\bibinfo{author}{Wang, H.} \emph{et~al.}
\newblock \bibinfo{journal}{\bibinfo{title}{Sub-gbps key rate four-state
  continuous-variable quantum key distribution within metropolitan area}}.
\newblock {\emph{\JournalTitle{Communications Physics}}}
  \textbf{\bibinfo{volume}{5}}, \bibinfo{pages}{162} (\bibinfo{year}{2022}).

\bibitem{grunenfelder2023fast}
\bibinfo{author}{Gr{\"u}nenfelder, F.} \emph{et~al.}
\newblock \bibinfo{journal}{\bibinfo{title}{Fast single-photon detectors and
  real-time key distillation enable high secret-key-rate quantum key
  distribution systems}}.
\newblock {\emph{\JournalTitle{Nature Photonics}}}
  \textbf{\bibinfo{volume}{17}}, \bibinfo{pages}{422--426}
  (\bibinfo{year}{2023}).

\bibitem{li2023high}
\bibinfo{author}{Li, W.} \emph{et~al.}
\newblock \bibinfo{journal}{\bibinfo{title}{High-rate quantum key distribution
  exceeding 110 mb s--1}}.
\newblock {\emph{\JournalTitle{Nature photonics}}}
  \textbf{\bibinfo{volume}{17}}, \bibinfo{pages}{416--421}
  (\bibinfo{year}{2023}).

\bibitem{hajomer2024cv}
\bibinfo{author}{Hajomer, A.~A.} \emph{et~al.}
\newblock \bibinfo{journal}{\bibinfo{title}{Continuous-variable quantum key
  distribution at 10 gbaud using an integrated photonic-electronic receiver}}.
\newblock {\emph{\JournalTitle{Optica}}} \textbf{\bibinfo{volume}{11}},
  \bibinfo{pages}{1197--1204} (\bibinfo{year}{2024}).

\bibitem{ji2024gbps}
\bibinfo{author}{Ji, F.}, \bibinfo{author}{Huang, P.}, \bibinfo{author}{Wang,
  T.}, \bibinfo{author}{Jiang, X.} \& \bibinfo{author}{Zeng, G.}
\newblock \bibinfo{journal}{\bibinfo{title}{Gbps key rate
  passive-state-preparation continuous-variable quantum key distribution within
  an access-network area}}.
\newblock {\emph{\JournalTitle{Photonics Research}}}
  \textbf{\bibinfo{volume}{12}}, \bibinfo{pages}{1485--1493}
  (\bibinfo{year}{2024}).

\bibitem{wang2025high}
\bibinfo{author}{Wang, H.} \emph{et~al.}
\newblock \bibinfo{journal}{\bibinfo{title}{High-rate continuous-variable
  quantum key distribution over 100 km fiber with composable security}}.
\newblock {\emph{\JournalTitle{Optica}}} \textbf{\bibinfo{volume}{12}},
  \bibinfo{pages}{1657--1667} (\bibinfo{year}{2025}).

\bibitem{hajomer2025chip}
\bibinfo{author}{Hajomer, A.~A.} \emph{et~al.}
\newblock \bibinfo{journal}{\bibinfo{title}{Chip-based 16 gbaud
  continuous-variable quantum key distribution}}.
\newblock {\emph{\JournalTitle{arXiv preprint arXiv:2504.09308}}}
  (\bibinfo{year}{2025}).

\bibitem{grosshans2002continuous}
\bibinfo{author}{Grosshans, F.} \& \bibinfo{author}{Grangier, P.}
\newblock \bibinfo{journal}{\bibinfo{title}{Continuous variable quantum
  cryptography using coherent states}}.
\newblock {\emph{\JournalTitle{Physical review letters}}}
  \textbf{\bibinfo{volume}{88}}, \bibinfo{pages}{057902}
  (\bibinfo{year}{2002}).

\bibitem{wang2005beating}
\bibinfo{author}{Wang, X.-B.}
\newblock \bibinfo{journal}{\bibinfo{title}{Beating the photon-number-splitting
  attack in practical quantum cryptography}}.
\newblock {\emph{\JournalTitle{Physical review letters}}}
  \textbf{\bibinfo{volume}{94}}, \bibinfo{pages}{230503}
  (\bibinfo{year}{2005}).

\bibitem{peng2007experimental}
\bibinfo{author}{Peng, C.-Z.} \emph{et~al.}
\newblock \bibinfo{journal}{\bibinfo{title}{Experimental long-distance
  decoy-state quantum key distribution based on polarization encoding}}.
\newblock {\emph{\JournalTitle{Physical review letters}}}
  \textbf{\bibinfo{volume}{98}}, \bibinfo{pages}{010505}
  (\bibinfo{year}{2007}).

\bibitem{bennett2014quantum}
\bibinfo{author}{Bennett, C.~H.} \& \bibinfo{author}{Brassard, G.}
\newblock \bibinfo{journal}{\bibinfo{title}{Quantum cryptography: Public key
  distribution and coin tossing}}.
\newblock {\emph{\JournalTitle{Theoretical computer science}}}
  \textbf{\bibinfo{volume}{560}}, \bibinfo{pages}{7--11}
  (\bibinfo{year}{2014}).

\bibitem{qi2015generating}
\bibinfo{author}{Qi, B.}, \bibinfo{author}{Lougovski, P.},
  \bibinfo{author}{Pooser, R.}, \bibinfo{author}{Grice, W.} \&
  \bibinfo{author}{Bobrek, M.}
\newblock \bibinfo{journal}{\bibinfo{title}{Generating the local oscillator
  ''locally'' in continuous-variable quantum key distribution based on coherent
  detection}}.
\newblock {\emph{\JournalTitle{Physical Review X}}}
  \textbf{\bibinfo{volume}{5}}, \bibinfo{pages}{041009} (\bibinfo{year}{2015}).

\bibitem{huang2016long}
\bibinfo{author}{Huang, D.}, \bibinfo{author}{Huang, P.}, \bibinfo{author}{Lin,
  D.} \& \bibinfo{author}{Zeng, G.}
\newblock \bibinfo{journal}{\bibinfo{title}{Long-distance continuous-variable
  quantum key distribution by controlling excess noise}}.
\newblock {\emph{\JournalTitle{Scientific reports}}}
  \textbf{\bibinfo{volume}{6}}, \bibinfo{pages}{19201} (\bibinfo{year}{2016}).

\bibitem{liao2017satellite}
\bibinfo{author}{Liao, S.-K.} \emph{et~al.}
\newblock \bibinfo{journal}{\bibinfo{title}{Satellite-to-ground quantum key
  distribution}}.
\newblock {\emph{\JournalTitle{Nature}}} \textbf{\bibinfo{volume}{549}},
  \bibinfo{pages}{43--47} (\bibinfo{year}{2017}).

\bibitem{lucamarini2018overcoming}
\bibinfo{author}{Lucamarini, M.}, \bibinfo{author}{Yuan, Z.~L.},
  \bibinfo{author}{Dynes, J.~F.} \& \bibinfo{author}{Shields, A.~J.}
\newblock \bibinfo{journal}{\bibinfo{title}{Overcoming the rate--distance limit
  of quantum key distribution without quantum repeaters}}.
\newblock {\emph{\JournalTitle{Nature}}} \textbf{\bibinfo{volume}{557}},
  \bibinfo{pages}{400--403} (\bibinfo{year}{2018}).

\bibitem{yin2020entanglement}
\bibinfo{author}{Yin, J.} \emph{et~al.}
\newblock \bibinfo{journal}{\bibinfo{title}{Entanglement-based secure quantum
  cryptography over 1,120 kilometres}}.
\newblock {\emph{\JournalTitle{Nature}}} \textbf{\bibinfo{volume}{582}},
  \bibinfo{pages}{501--505} (\bibinfo{year}{2020}).

\bibitem{zhang2020long}
\bibinfo{author}{Zhang, Y.} \emph{et~al.}
\newblock \bibinfo{journal}{\bibinfo{title}{Long-distance continuous-variable
  quantum key distribution over 202.81 km of fiber}}.
\newblock {\emph{\JournalTitle{Physical review letters}}}
  \textbf{\bibinfo{volume}{125}}, \bibinfo{pages}{010502}
  (\bibinfo{year}{2020}).

\bibitem{wang2022twin}
\bibinfo{author}{Wang, S.} \emph{et~al.}
\newblock \bibinfo{journal}{\bibinfo{title}{Twin-field quantum key distribution
  over 830-km fibre}}.
\newblock {\emph{\JournalTitle{Nature photonics}}}
  \textbf{\bibinfo{volume}{16}}, \bibinfo{pages}{154--161}
  (\bibinfo{year}{2022}).

\bibitem{liu2023experimental}
\bibinfo{author}{Liu, Y.} \emph{et~al.}
\newblock \bibinfo{journal}{\bibinfo{title}{Experimental twin-field quantum key
  distribution over 1000 km fiber distance}}.
\newblock {\emph{\JournalTitle{Physical Review Letters}}}
  \textbf{\bibinfo{volume}{130}}, \bibinfo{pages}{210801}
  (\bibinfo{year}{2023}).

\bibitem{hajomer2024long}
\bibinfo{author}{Hajomer, A.~A.} \emph{et~al.}
\newblock \bibinfo{journal}{\bibinfo{title}{Long-distance continuous-variable
  quantum key distribution over 100-km fiber with local local oscillator}}.
\newblock {\emph{\JournalTitle{Science Advances}}}
  \textbf{\bibinfo{volume}{10}}, \bibinfo{pages}{eadi9474}
  (\bibinfo{year}{2024}).

\bibitem{townsend1997quantum}
\bibinfo{author}{Townsend, P.~D.}
\newblock \bibinfo{journal}{\bibinfo{title}{Quantum cryptography on multiuser
  optical fibre networks}}.
\newblock {\emph{\JournalTitle{Nature}}} \textbf{\bibinfo{volume}{385}},
  \bibinfo{pages}{47--49} (\bibinfo{year}{1997}).

\bibitem{choi2011quantum}
\bibinfo{author}{Choi, I.}, \bibinfo{author}{Young, R.~J.} \&
  \bibinfo{author}{Townsend, P.~D.}
\newblock \bibinfo{journal}{\bibinfo{title}{Quantum information to the home}}.
\newblock {\emph{\JournalTitle{New Journal of Physics}}}
  \textbf{\bibinfo{volume}{13}}, \bibinfo{pages}{063039}
  (\bibinfo{year}{2011}).

\bibitem{frohlich2013quantum}
\bibinfo{author}{Fr{\"o}hlich, B.} \emph{et~al.}
\newblock \bibinfo{journal}{\bibinfo{title}{A quantum access network}}.
\newblock {\emph{\JournalTitle{Nature}}} \textbf{\bibinfo{volume}{501}},
  \bibinfo{pages}{69--72} (\bibinfo{year}{2013}).

\bibitem{frohlich2015quantum}
\bibinfo{author}{Fr{\"o}hlich, B.} \emph{et~al.}
\newblock \bibinfo{journal}{\bibinfo{title}{Quantum secured gigabit optical
  access networks}}.
\newblock {\emph{\JournalTitle{Scientific reports}}}
  \textbf{\bibinfo{volume}{5}}, \bibinfo{pages}{18121} (\bibinfo{year}{2015}).

\bibitem{wengerowsky2018entanglement}
\bibinfo{author}{Wengerowsky, S.}, \bibinfo{author}{Joshi, S.~K.},
  \bibinfo{author}{Steinlechner, F.}, \bibinfo{author}{H{\"u}bel, H.} \&
  \bibinfo{author}{Ursin, R.}
\newblock \bibinfo{journal}{\bibinfo{title}{An entanglement-based
  wavelength-multiplexed quantum communication network}}.
\newblock {\emph{\JournalTitle{Nature}}} \textbf{\bibinfo{volume}{564}},
  \bibinfo{pages}{225--228} (\bibinfo{year}{2018}).

\bibitem{dynes2019cambridge}
\bibinfo{author}{Dynes, J.} \emph{et~al.}
\newblock \bibinfo{journal}{\bibinfo{title}{Cambridge quantum network}}.
\newblock {\emph{\JournalTitle{npj Quantum Information}}}
  \textbf{\bibinfo{volume}{5}}, \bibinfo{pages}{101} (\bibinfo{year}{2019}).

\bibitem{joshi2020trusted}
\bibinfo{author}{Joshi, S.~K.} \emph{et~al.}
\newblock \bibinfo{journal}{\bibinfo{title}{A trusted node--free eight-user
  metropolitan quantum communication network}}.
\newblock {\emph{\JournalTitle{Science advances}}}
  \textbf{\bibinfo{volume}{6}}, \bibinfo{pages}{eaba0959}
  (\bibinfo{year}{2020}).

\bibitem{chen2021implementation}
\bibinfo{author}{Chen, T.-Y.} \emph{et~al.}
\newblock \bibinfo{journal}{\bibinfo{title}{Implementation of a 46-node quantum
  metropolitan area network}}.
\newblock {\emph{\JournalTitle{npj Quantum Information}}}
  \textbf{\bibinfo{volume}{7}}, \bibinfo{pages}{134} (\bibinfo{year}{2021}).

\bibitem{chen2021integrated}
\bibinfo{author}{Chen, Y.-A.} \emph{et~al.}
\newblock \bibinfo{journal}{\bibinfo{title}{An integrated space-to-ground
  quantum communication network over 4,600 kilometres}}.
\newblock {\emph{\JournalTitle{Nature}}} \textbf{\bibinfo{volume}{589}},
  \bibinfo{pages}{214--219} (\bibinfo{year}{2021}).

\bibitem{huang2021realizing}
\bibinfo{author}{Huang, Y.} \emph{et~al.}
\newblock \bibinfo{journal}{\bibinfo{title}{Realizing a downstream-access
  network using continuous-variable quantum key distribution}}.
\newblock {\emph{\JournalTitle{Physical Review Applied}}}
  \textbf{\bibinfo{volume}{16}}, \bibinfo{pages}{064051}
  (\bibinfo{year}{2021}).

\bibitem{wang2021practical}
\bibinfo{author}{Wang, B.-X.} \emph{et~al.}
\newblock \bibinfo{journal}{\bibinfo{title}{Practical quantum access network
  over a 10 gbit/s ethernet passive optical network}}.
\newblock {\emph{\JournalTitle{Optics Express}}} \textbf{\bibinfo{volume}{29}},
  \bibinfo{pages}{38582--38590} (\bibinfo{year}{2021}).

\bibitem{fan2022robust}
\bibinfo{author}{Fan-Yuan, G.-J.} \emph{et~al.}
\newblock \bibinfo{journal}{\bibinfo{title}{Robust and adaptable quantum key
  distribution network without trusted nodes}}.
\newblock {\emph{\JournalTitle{Optica}}} \textbf{\bibinfo{volume}{9}},
  \bibinfo{pages}{812--823} (\bibinfo{year}{2022}).

\bibitem{mandil2023quantum}
\bibinfo{author}{Mandil, R.}, \bibinfo{author}{DiAdamo, S.},
  \bibinfo{author}{Qi, B.} \& \bibinfo{author}{Shabani, A.}
\newblock \bibinfo{journal}{\bibinfo{title}{Quantum key distribution in a
  packet-switched network}}.
\newblock {\emph{\JournalTitle{npj Quantum Information}}}
  \textbf{\bibinfo{volume}{9}}, \bibinfo{pages}{85} (\bibinfo{year}{2023}).

\bibitem{wang2023experimental}
\bibinfo{author}{Wang, X.} \emph{et~al.}
\newblock \bibinfo{journal}{\bibinfo{title}{Experimental upstream transmission
  of continuous variable quantum key distribution access network}}.
\newblock {\emph{\JournalTitle{Optics Letters}}} \textbf{\bibinfo{volume}{48}},
  \bibinfo{pages}{3327--3330} (\bibinfo{year}{2023}).

\bibitem{xu2023round}
\bibinfo{author}{Xu, Y.}, \bibinfo{author}{Wang, T.}, \bibinfo{author}{Zhao,
  H.}, \bibinfo{author}{Huang, P.} \& \bibinfo{author}{Zeng, G.}
\newblock \bibinfo{journal}{\bibinfo{title}{Round-trip multi-band quantum
  access network}}.
\newblock {\emph{\JournalTitle{Photonics Research}}}
  \textbf{\bibinfo{volume}{11}}, \bibinfo{pages}{1449--1464}
  (\bibinfo{year}{2023}).

\bibitem{hajomer2024continuous}
\bibinfo{author}{Hajomer, A.~A.} \emph{et~al.}
\newblock \bibinfo{journal}{\bibinfo{title}{Continuous-variable quantum passive
  optical network}}.
\newblock {\emph{\JournalTitle{Light: Science \& Applications}}}
  \textbf{\bibinfo{volume}{13}}, \bibinfo{pages}{291} (\bibinfo{year}{2024}).

\bibitem{huang2024cost}
\bibinfo{author}{Huang, C.} \emph{et~al.}
\newblock \bibinfo{journal}{\bibinfo{title}{A cost-efficient quantum access
  network with qubit-based synchronization}}.
\newblock {\emph{\JournalTitle{Science China Physics, Mechanics \& Astronomy}}}
  \textbf{\bibinfo{volume}{67}}, \bibinfo{pages}{240312}
  (\bibinfo{year}{2024}).

\bibitem{li2024experimental}
\bibinfo{author}{Li, Z.}, \bibinfo{author}{Wang, X.}, \bibinfo{author}{Qi, D.},
  \bibinfo{author}{Chen, Z.} \& \bibinfo{author}{Yu, S.}
\newblock \bibinfo{journal}{\bibinfo{title}{Experimental implementation of
  four-user downstream access network continuous-variable quantum key
  distribution}}.
\newblock {\emph{\JournalTitle{Journal of Lightwave Technology}}}
  (\bibinfo{year}{2024}).

\bibitem{liu2024integrated}
\bibinfo{author}{Liu, S.} \emph{et~al.}
\newblock \bibinfo{journal}{\bibinfo{title}{Integrated quantum communication
  network and vibration sensing in optical fibers}}.
\newblock {\emph{\JournalTitle{Optica}}} \textbf{\bibinfo{volume}{11}},
  \bibinfo{pages}{1762--1772} (\bibinfo{year}{2024}).

\bibitem{qi2024experimental}
\bibinfo{author}{Qi, D.} \emph{et~al.}
\newblock \bibinfo{journal}{\bibinfo{title}{Experimental demonstration of a
  quantum downstream access network in continuous variable quantum key
  distribution with a local local oscillator}}.
\newblock {\emph{\JournalTitle{Photonics Research}}}
  \textbf{\bibinfo{volume}{12}}, \bibinfo{pages}{1262--1273}
  (\bibinfo{year}{2024}).

\bibitem{xu2024integrated}
\bibinfo{author}{Xu, Y.}, \bibinfo{author}{Wang, T.}, \bibinfo{author}{Huang,
  P.} \& \bibinfo{author}{Zeng, G.}
\newblock \bibinfo{journal}{\bibinfo{title}{Integrated distributed sensing and
  quantum communication networks}}.
\newblock {\emph{\JournalTitle{Research}}} \textbf{\bibinfo{volume}{7}},
  \bibinfo{pages}{0416} (\bibinfo{year}{2024}).

\bibitem{pan2025high}
\bibinfo{author}{Pan, Y.} \emph{et~al.}
\newblock \bibinfo{journal}{\bibinfo{title}{High-rate 16-node quantum access
  network based on a passive optical network}}.
\newblock {\emph{\JournalTitle{Optica}}} \textbf{\bibinfo{volume}{12}},
  \bibinfo{pages}{953--960} (\bibinfo{year}{2025}).

\bibitem{xu2025ofdm}
\bibinfo{author}{Xu, Y.} \emph{et~al.}
\newblock \bibinfo{journal}{\bibinfo{title}{Ofdm-based quantum key distribution
  access network reaching nyquist limits}}.
\newblock {\emph{\JournalTitle{Optica}}} \textbf{\bibinfo{volume}{12}},
  \bibinfo{pages}{1668--1680} (\bibinfo{year}{2025}).

\bibitem{xu2026polychromatic}
\bibinfo{author}{Xu, Y.} \emph{et~al.}
\newblock \bibinfo{journal}{\bibinfo{title}{Polychromatic continuous-variable
  quantum communication network enabled by optical frequency combs}}.
\newblock {\emph{\JournalTitle{npj Quantum Information}}}
  (\bibinfo{year}{2026}).

\bibitem{kramer2002ethernet}
\bibinfo{author}{Kramer, G.} \& \bibinfo{author}{Pesavento, G.}
\newblock \bibinfo{journal}{\bibinfo{title}{Ethernet passive optical network
  (epon): Building a next-generation optical access network}}.
\newblock {\emph{\JournalTitle{IEEE Communications magazine}}}
  \textbf{\bibinfo{volume}{40}}, \bibinfo{pages}{66--73}
  (\bibinfo{year}{2002}).

\bibitem{banerjee2005wavelength}
\bibinfo{author}{Banerjee, A.} \emph{et~al.}
\newblock \bibinfo{journal}{\bibinfo{title}{Wavelength-division-multiplexed
  passive optical network (wdm-pon) technologies for broadband access: a
  review}}.
\newblock {\emph{\JournalTitle{Journal of optical networking}}}
  \textbf{\bibinfo{volume}{4}}, \bibinfo{pages}{737--758}
  (\bibinfo{year}{2005}).

\bibitem{mcgarry2006wdm}
\bibinfo{author}{McGarry, M.~P.}, \bibinfo{author}{Reisslein, M.} \&
  \bibinfo{author}{Maier, M.}
\newblock \bibinfo{journal}{\bibinfo{title}{Wdm ethernet passive optical
  networks}}.
\newblock {\emph{\JournalTitle{IEEE Communications Magazine}}}
  \textbf{\bibinfo{volume}{44}}, \bibinfo{pages}{15--22}
  (\bibinfo{year}{2006}).

\bibitem{lam2011passive}
\bibinfo{author}{Lam, C.~F.}
\newblock \emph{\bibinfo{title}{Passive optical networks: principles and
  practice}} (\bibinfo{publisher}{Elsevier}, \bibinfo{year}{2011}).

\bibitem{abbas2016next}
\bibinfo{author}{Abbas, H.~S.} \& \bibinfo{author}{Gregory, M.~A.}
\newblock \bibinfo{journal}{\bibinfo{title}{The next generation of passive
  optical networks: A review}}.
\newblock {\emph{\JournalTitle{Journal of network and computer applications}}}
  \textbf{\bibinfo{volume}{67}}, \bibinfo{pages}{53--74}
  (\bibinfo{year}{2016}).

\bibitem{wey2018passive}
\bibinfo{author}{Wey, J.~S.} \& \bibinfo{author}{Zhang, J.}
\newblock \bibinfo{journal}{\bibinfo{title}{Passive optical networks for 5g
  transport: technology and standards}}.
\newblock {\emph{\JournalTitle{Journal of Lightwave Technology}}}
  \textbf{\bibinfo{volume}{37}}, \bibinfo{pages}{2830--2837}
  (\bibinfo{year}{2018}).

\bibitem{qi2018passive}
\bibinfo{author}{Qi, B.}, \bibinfo{author}{Evans, P.~G.} \&
  \bibinfo{author}{Grice, W.~P.}
\newblock \bibinfo{journal}{\bibinfo{title}{Passive state preparation in the
  gaussian-modulated coherent-states quantum key distribution}}.
\newblock {\emph{\JournalTitle{Physical Review A}}}
  \textbf{\bibinfo{volume}{97}}, \bibinfo{pages}{012317}
  (\bibinfo{year}{2018}).

\bibitem{qi2020experimental}
\bibinfo{author}{Qi, B.} \emph{et~al.}
\newblock \bibinfo{journal}{\bibinfo{title}{Experimental passive-state
  preparation for continuous-variable quantum communications}}.
\newblock {\emph{\JournalTitle{Physical Review Applied}}}
  \textbf{\bibinfo{volume}{13}}, \bibinfo{pages}{054065}
  (\bibinfo{year}{2020}).

\bibitem{huang2021experimental}
\bibinfo{author}{Huang, P.} \emph{et~al.}
\newblock \bibinfo{journal}{\bibinfo{title}{Experimental continuous-variable
  quantum key distribution using a thermal source}}.
\newblock {\emph{\JournalTitle{New Journal of Physics}}}
  \textbf{\bibinfo{volume}{23}}, \bibinfo{pages}{113028}
  (\bibinfo{year}{2021}).

\bibitem{wu2021passive}
\bibinfo{author}{Wu, X.}, \bibinfo{author}{Wang, Y.}, \bibinfo{author}{Guo,
  Y.}, \bibinfo{author}{Zhong, H.} \& \bibinfo{author}{Huang, D.}
\newblock \bibinfo{journal}{\bibinfo{title}{Passive continuous-variable quantum
  key distribution using a locally generated local oscillator}}.
\newblock {\emph{\JournalTitle{Physical Review A}}}
  \textbf{\bibinfo{volume}{103}}, \bibinfo{pages}{032604}
  (\bibinfo{year}{2021}).

\bibitem{zhang2023experimental}
\bibinfo{author}{Zhang, M.}, \bibinfo{author}{Huang, P.},
  \bibinfo{author}{Wang, P.}, \bibinfo{author}{Wei, S.} \&
  \bibinfo{author}{Zeng, G.}
\newblock \bibinfo{journal}{\bibinfo{title}{Experimental free-space
  continuous-variable quantum key distribution with thermal source}}.
\newblock {\emph{\JournalTitle{Optics Letters}}} \textbf{\bibinfo{volume}{48}},
  \bibinfo{pages}{1184--1187} (\bibinfo{year}{2023}).

\bibitem{yin2025all}
\bibinfo{author}{Yin, H.} \emph{et~al.}
\newblock \bibinfo{journal}{\bibinfo{title}{All-day free-space quantum key
  distribution with thermal source towards quantum secure communications for
  unmanned vehicles}}.
\newblock {\emph{\JournalTitle{npj Quantum Information}}}
  \textbf{\bibinfo{volume}{11}}, \bibinfo{pages}{134} (\bibinfo{year}{2025}).

\bibitem{raymer1989temporal}
\bibinfo{author}{Raymer, M.}, \bibinfo{author}{Li, Z.} \&
  \bibinfo{author}{Walmsley, I.}
\newblock \bibinfo{journal}{\bibinfo{title}{Temporal quantum fluctuations in
  stimulated raman scattering: Coherent-modes description}}.
\newblock {\emph{\JournalTitle{Physical review letters}}}
  \textbf{\bibinfo{volume}{63}}, \bibinfo{pages}{1586} (\bibinfo{year}{1989}).

\bibitem{blow1990continuum}
\bibinfo{author}{Blow, K.}, \bibinfo{author}{Loudon, R.},
  \bibinfo{author}{Phoenix, S.~J.} \& \bibinfo{author}{Shepherd, T.}
\newblock \bibinfo{journal}{\bibinfo{title}{Continuum fields in quantum
  optics}}.
\newblock {\emph{\JournalTitle{Physical Review A}}}
  \textbf{\bibinfo{volume}{42}}, \bibinfo{pages}{4102} (\bibinfo{year}{1990}).

\bibitem{fabre2020modes}
\bibinfo{author}{Fabre, C.} \& \bibinfo{author}{Treps, N.}
\newblock \bibinfo{journal}{\bibinfo{title}{Modes and states in quantum
  optics}}.
\newblock {\emph{\JournalTitle{Reviews of Modern Physics}}}
  \textbf{\bibinfo{volume}{92}}, \bibinfo{pages}{035005}
  (\bibinfo{year}{2020}).

\bibitem{raymer2020temporal}
\bibinfo{author}{Raymer, M.~G.} \& \bibinfo{author}{Walmsley, I.~A.}
\newblock \bibinfo{journal}{\bibinfo{title}{Temporal modes in quantum optics:
  then and now}}.
\newblock {\emph{\JournalTitle{Physica Scripta}}}
  \textbf{\bibinfo{volume}{95}}, \bibinfo{pages}{064002}
  (\bibinfo{year}{2020}).

\bibitem{chen2023continuous}
\bibinfo{author}{Chen, Z.}, \bibinfo{author}{Wang, X.}, \bibinfo{author}{Yu,
  S.}, \bibinfo{author}{Li, Z.} \& \bibinfo{author}{Guo, H.}
\newblock \bibinfo{journal}{\bibinfo{title}{Continuous-mode quantum key
  distribution with digital signal processing}}.
\newblock {\emph{\JournalTitle{npj Quantum Information}}}
  \textbf{\bibinfo{volume}{9}}, \bibinfo{pages}{28} (\bibinfo{year}{2023}).

\bibitem{sun2025analyzing}
\bibinfo{author}{Sun, Y.}, \bibinfo{author}{Chen, Z.}, \bibinfo{author}{Wang,
  X.}, \bibinfo{author}{Yu, S.} \& \bibinfo{author}{Guo, H.}
\newblock \bibinfo{journal}{\bibinfo{title}{Analyzing the performance of cv-mdi
  qkd under continuous-mode scenarios}}.
\newblock {\emph{\JournalTitle{Physical Review Applied}}}
  \textbf{\bibinfo{volume}{23}}, \bibinfo{pages}{014056}
  (\bibinfo{year}{2025}).

\bibitem{xu2024robust}
\bibinfo{author}{Xu, Y.} \emph{et~al.}
\newblock \bibinfo{journal}{\bibinfo{title}{Robust continuous-variable quantum
  key distribution in the finite-size regime}}.
\newblock {\emph{\JournalTitle{Photonics Research}}}
  \textbf{\bibinfo{volume}{12}}, \bibinfo{pages}{2549--2558}
  (\bibinfo{year}{2024}).

\bibitem{liao2025high}
\bibinfo{author}{Liao, X.} \emph{et~al.}
\newblock \bibinfo{journal}{\bibinfo{title}{High-rate self-referenced
  continuous-variable quantum key distribution over a high-loss free-space
  channel}}.
\newblock {\emph{\JournalTitle{Photonics Research}}}
  \textbf{\bibinfo{volume}{13}}, \bibinfo{pages}{2603--2617}
  (\bibinfo{year}{2025}).

\bibitem{droste2013optical}
\bibinfo{author}{Droste, S.} \emph{et~al.}
\newblock \bibinfo{journal}{\bibinfo{title}{Optical-frequency transfer over a
  single-span 1840 km fiber link}}.
\newblock {\emph{\JournalTitle{Physical review letters}}}
  \textbf{\bibinfo{volume}{111}}, \bibinfo{pages}{110801}
  (\bibinfo{year}{2013}).

\bibitem{chen2024dual}
\bibinfo{author}{Chen, Z.} \emph{et~al.}
\newblock \bibinfo{journal}{\bibinfo{title}{Dual-comb-enhanced microwave clock
  synchronization over commercial fiber}}.
\newblock {\emph{\JournalTitle{Optica}}} \textbf{\bibinfo{volume}{11}},
  \bibinfo{pages}{1268--1276} (\bibinfo{year}{2024}).

\bibitem{scully1997quantum}
\bibinfo{author}{Scully, M.~O.} \& \bibinfo{author}{Zubairy, M.~S.}
\newblock \emph{\bibinfo{title}{Quantum optics}} (\bibinfo{publisher}{Cambridge
  university press}, \bibinfo{year}{1997}).

\bibitem{leverrier2008multidimensional}
\bibinfo{author}{Leverrier, A.}, \bibinfo{author}{All{\'e}aume, R.},
  \bibinfo{author}{Boutros, J.}, \bibinfo{author}{Z{\'e}mor, G.} \&
  \bibinfo{author}{Grangier, P.}
\newblock \bibinfo{journal}{\bibinfo{title}{Multidimensional reconciliation for
  a continuous-variable quantum key distribution}}.
\newblock {\emph{\JournalTitle{Physical Review A}}}
  \textbf{\bibinfo{volume}{77}}, \bibinfo{pages}{042325}
  (\bibinfo{year}{2008}).

\bibitem{laudenbach2018continuous}
\bibinfo{author}{Laudenbach, F.} \emph{et~al.}
\newblock \bibinfo{journal}{\bibinfo{title}{Continuous-variable quantum key
  distribution with gaussian modulation----the theory of practical
  implementations}}.
\newblock {\emph{\JournalTitle{Advanced Quantum Technologies}}}
  \textbf{\bibinfo{volume}{1}}, \bibinfo{pages}{1800011}
  (\bibinfo{year}{2018}).

\bibitem{leverrier2010finite}
\bibinfo{author}{Leverrier, A.}, \bibinfo{author}{Grosshans, F.} \&
  \bibinfo{author}{Grangier, P.}
\newblock \bibinfo{journal}{\bibinfo{title}{Finite-size analysis of a
  continuous-variable quantum key distribution}}.
\newblock {\emph{\JournalTitle{Physical Review A}}}
  \textbf{\bibinfo{volume}{81}}, \bibinfo{pages}{062343}
  (\bibinfo{year}{2010}).

\bibitem{leverrier2015composable}
\bibinfo{author}{Leverrier, A.}
\newblock \bibinfo{journal}{\bibinfo{title}{Composable security proof for
  continuous-variable quantum key distribution with coherent states}}.
\newblock {\emph{\JournalTitle{Physical review letters}}}
  \textbf{\bibinfo{volume}{114}}, \bibinfo{pages}{070501}
  (\bibinfo{year}{2015}).

\bibitem{pirandola2017fundamental}
\bibinfo{author}{Pirandola, S.}, \bibinfo{author}{Laurenza, R.},
  \bibinfo{author}{Ottaviani, C.} \& \bibinfo{author}{Banchi, L.}
\newblock \bibinfo{journal}{\bibinfo{title}{Fundamental limits of repeaterless
  quantum communications}}.
\newblock {\emph{\JournalTitle{Nature communications}}}
  \textbf{\bibinfo{volume}{8}}, \bibinfo{pages}{1--15} (\bibinfo{year}{2017}).

\bibitem{pirandola2019bounds}
\bibinfo{author}{Pirandola, S.}
\newblock \bibinfo{journal}{\bibinfo{title}{Bounds for multi-end communication
  over quantum networks}}.
\newblock {\emph{\JournalTitle{Quantum Science and Technology}}}
  \textbf{\bibinfo{volume}{4}}, \bibinfo{pages}{045006} (\bibinfo{year}{2019}).

\bibitem{pirandola2019end}
\bibinfo{author}{Pirandola, S.}
\newblock \bibinfo{journal}{\bibinfo{title}{End-to-end capacities of a quantum
  communication network}}.
\newblock {\emph{\JournalTitle{Communications Physics}}}
  \textbf{\bibinfo{volume}{2}}, \bibinfo{pages}{51} (\bibinfo{year}{2019}).

\bibitem{das2021universal}
\bibinfo{author}{Das, S.}, \bibinfo{author}{B{\"a}uml, S.},
  \bibinfo{author}{Winczewski, M.} \& \bibinfo{author}{Horodecki, K.}
\newblock \bibinfo{journal}{\bibinfo{title}{Universal limitations on quantum
  key distribution over a network}}.
\newblock {\emph{\JournalTitle{Physical Review X}}}
  \textbf{\bibinfo{volume}{11}}, \bibinfo{pages}{041016}
  (\bibinfo{year}{2021}).

\bibitem{zhang2025effective}
\bibinfo{author}{Zhang, K.} \emph{et~al.}
\newblock \bibinfo{journal}{\bibinfo{title}{Effective rate-adaptive
  reconciliation for cv-qkd using qc-met-ldpc codes}}.
\newblock {\emph{\JournalTitle{Science China Information Sciences}}}
  \textbf{\bibinfo{volume}{68}}, \bibinfo{pages}{180510}
  (\bibinfo{year}{2025}).

\bibitem{hajomer2022modulation}
\bibinfo{author}{Hajomer, A.~A.} \emph{et~al.}
\newblock \bibinfo{journal}{\bibinfo{title}{Modulation leakage-free
  continuous-variable quantum key distribution}}.
\newblock {\emph{\JournalTitle{npj Quantum Information}}}
  \textbf{\bibinfo{volume}{8}}, \bibinfo{pages}{136} (\bibinfo{year}{2022}).

\end{thebibliography}

\end{document}